\documentclass[nojss]{jss}

\usepackage{amsmath}
\usepackage{caption}
\usepackage{subcaption}

\author{Isabella Gollini\\University of Bristol, UK \And 
        Binbin Lu\\ Wuhan University, China\AND
        Martin Charlton\\ NUI Maynooth, Ireland\And
       Christopher Brunsdon\\NUI Maynooth, Ireland\And 
       Paul Harris\\ Rothamsted Research, UK}
\title{\pkg{GWmodel}: an \proglang{R} Package for Exploring Spatial Heterogeneity using Geographically Weighted Models}

\Plainauthor{Isabella Gollini, Binbin Lu, Martin Charlton, Christopher Brunsdon, Paul Harris} 
\Plaintitle{GWmodel: an R Package for Exploring Spatial Heterogeneity 
} 
\Shorttitle{\pkg{GWmodel}: Geographically Weighted Models} 

\Abstract{
Spatial statistics is a growing discipline providing important analytical techniques in a wide range of disciplines in the natural and social sciences. In the  \proglang{R} package \pkg{GWmodel}, we present techniques from a particular branch of spatial statistics, termed geographically weighted (GW) models. GW models suit situations when data are not described well by some global model, but where there are spatial regions where a suitably localised calibration provides a better description. The approach uses a moving window weighting technique, where localised models are found at target locations. Outputs are mapped to provide a useful exploratory tool into the nature of the data spatial heterogeneity. Currently, \pkg{GWmodel} includes functions for: GW summary statistics, GW principal components analysis, GW regression, and GW discriminant analysis; some of which are provided in basic and robust forms.
}
\Plainkeywords{geographically weighted regression, geographically weighted principal components analysis, spatial prediction, R package} 

\Address{
Isabella Gollini\\
Department of Civil Engineering\\
University of Bristol\\
Queen's Building\\
Bristol BS8 1TR, UK\\
E-mail: \email{isabella.gollini@bristol.ac.uk}\\
URL: \url{http://www.bristol.ac.uk/engineering/people/isabella-gollini/}
}

\begin{document}

\section[Introduction]{Introduction}\label{sec.intro}
Spatial statistics provides important analytical techniques for a wide range of disciplines in the natural and social sciences, where (often large) spatial data sets are routinely collected.  Here we present techniques from a particular branch of non-stationary spatial statistics, termed geographically weighted (GW) models.  GW models suit situations when spatial data are not described well by some universal or global model, but where there are spatial regions where a suitably localised model calibration provides a better description.  The approach uses a moving window weighting technique, where localised models are found at target locations.  Here, for an individual model at some target location, we weight all neighbouring observations according to some distance-decay kernel function and then locally apply the model to this weighted data.  The size of the window over which this localised model might apply is controlled by the bandwidth.  Small bandwidths lead to more rapid spatial variation in the results while large bandwidths yield results increasingly close to the universal model solution.  When there exists some objective function (e.g., the model can predict), a bandwidth can be found optimally, using cross-validation and related approaches.

The GW modelling paradigm has evolved to encompass many techniques; techniques that are applicable when a certain heterogeneity or non-stationarity is suspected in the study's spatial process.  Commonly, outputs or parameters of the GW model are mapped to provide a useful exploratory tool, which can often precede (and direct) a more traditional or sophisticated statistical analysis.  Subsequent analyses can be non-spatial or spatial, where the latter can incorporate stationary or non-stationary decisions.  Notable GW models include: GW summary statistics \citep{BrFoCh:02a}; GW principal components analysis (GW PCA) \citep{FoBrCh:02a,lloy:10a,gwpca11}; GW regression \citep{BrFoCh:96a,BrFoCh:98,BrFoCh:99,LeMeZh:00a,whe07}; GW generalised linear models \citep{FoBrCh:02a,nafobrch05}; GW discriminant analysis \citep{BrFoCh:07}; GW variograms \citep{hachfo10}; GW regression kriging hybrids \citep{haju11} and GW visualisation techniques \citep{dybr07}.

Many of these GW models are included in the \proglang{R} package \pkg{GWmodel} that we describe in this paper.  Those that are not, will be incorporated at a later date.  For the GW models that are included, there is a clear emphasis on data exploration.  Notably, \pkg{GWmodel} provides functions to conduct: (i) a GW PCA; (ii) GW regression with a local ridge compensation (for addressing local collinearity); (iii) mixed GW regression; (iv) heteroskedastic GW regression; (v) a GW discriminant analysis; (vi) robust and outlier-resistant GW modelling; (vii) Monte Carlo significance tests for non-stationarity; and (viii) GW modelling with a wide selection of distance metric and kernel weighting options.  These functions extend and enhance functions for: (a) GW summary statistics; (b) basic GW regression; and (c) GW generalised linear models - GW models that are also found in the \pkg{spgwr} \proglang{R} package \citep{spgwr}.  In this respect, \pkg{GWmodel} provides a more extensive set of GW modelling tools, within a single coherent framework (\pkg{GWmodel} similarly extends or complements the \pkg{gwrr} \proglang{R} package \citep{gwrr} with respect to GW regression and local collinearity issues).  \pkg{GWmodel} also provides an advanced alternative to various executable software packages that have a focus on GW regression - such as \proglang{GW regression v3.0} \citep{gwr3}; the \proglang{ArcGIS} GW regression tool in the Spatial Statistics Toolbox \citep{arcgis}; \proglang{SAM} for GW regression applications in macroecology \citep{sam10}; and \proglang{SpaceStat} for GW regression applications in health \citep{spacestat}.

Noting that it is not feasible to describe in detail all of the available functions in \pkg{GWmodel}, our paper has a robust theme and is structured as follows.  Section~\ref{sec.dataset} describes the example data sets that are available in \pkg{GWmodel}.  Section~\ref{sec.distbw} describes the various distance metric and kernel weighting options.  Section~\ref{sec.gwss} describes modelling with basic and robust GW summary statistics.  Section~\ref{sec.gwpca} describes modelling with basic and robust GW PCA.  Section~\ref{sec.gwr} describes modelling with basic and robust GW regression.  Section~\ref{sec.multicoll} describes ways to address local collinearity issues when modelling with GW regression.  Section~\ref{sec.gwpred} describes how to use GW regression as a spatial predictor.  Section~\ref{sec.gwcomp} relates the functions of \pkg {GWmodel} to those found in the \pkg{spgwr}, \pkg{gwrr} and \pkg{McSpatial} \citep{McSpatial} \proglang{R} packages. Section~\ref{sec.disc} concludes this work and indicates future work.

\section{Data sets} \label{sec.dataset}
The \pkg{GWmodel} package comes with five example data sets, these are: (i) \code{Georgia}, (ii) \code{LondonHP}, (iii) \code{USelect}, (iv) \code{DubVoter}, and (v) \code{EWHP}.  The \code{Georgia} data consists of selected 1990 US census variables (with $n=159$) for counties in the US state of Georgia; and is fully described in \cite{FoBrCh:02a}.  This data has been routinely used in a GW regression context for linking educational attainment with various contextual social variables \citep[see also][]{gri08}.  The data set is also available in the \proglang{GW regression 3.0} executable software package \citep{gwr3} and the \pkg{spgwr} \proglang{R} package.

The \code{LondonHP} data is a house price data set for London, England.  This data set (with $n=372$) is sampled from a 2001 house price data set, provided by the Nationwide Building Society of the UK and is combined with various hedonic contextual variables \citep{FoBrCh:02a}.  The hedonic data reflect structural characteristics of the property, property construction time, property type and local household income conditions.  Studies in house price markets with respect to modelling hedonic relationships has been a common application of GW regression  \citep[e.g., ][]{KeThRo06,BiMuDa07,PaFeFa08}.

The \code{USelect} data consists of the results of the 2004 US presidential election at the county level, together with five census variables (with $n=3111$).  The data is a subset of that provided in \citep{Robin:13}. \code{USelect} is similar to that used for the visualisation of GW discriminant analysis outputs in \citep{FoDem:13}; the only difference is that we specify the categorical, election results variable with three classes (instead of two): (a) Bush winner, (b) Kerry winner and (c) Borderline (for marginal winning results).  

For this article's presentation of GW models, we use as case studies, the \code{DubVoter} and \code{EWHP} data sets.  The \code{DubVoter} data (with $n=322$) is the main study data set and is used throughout Sections~\ref{sec.gwss} to \ref{sec.multicoll}, where key GW models are presented.  This data is composed of nine percentage variables\footnote{Observe that none of the \code{DubVoter} variables constitute a closed system (i.e., the full array of values sum to 100) and as such, we do not need to transform the data prior to a (univariate or multivariate) GW model calibration.}, measuring: (1) voter turnout in the Irish 2004 D\'ail elections and (2) eight characteristics of social structure (census data); for 322 Electoral Divisions (EDs) of Greater Dublin. \cite{kaetal06} modelled this data using GW regression; with voter turnout (\code{GenEl2004}) the dependent variable (i.e., the percentage of the population in each ED who voted in the election).  The eight independent variables measure the percentage of the population in each ED, with respect to:
\begin{itemize}
\item[A.]	one year migrants (i.e., moved to a different address one year ago) (\code{DiffAdd});
\item[B.]	local authority renters (\code{LARent});
\item[C.]	social class one (high social class) (\code{SC1});
\item[D.]	unemployed (\code{Unempl});
\item[E.]	without any formal educational (\code{LowEduc});
\item[F.]	age group 18-24 (\code{Age18_24});
\item[G.]	age group 25-44 (\code{Age25_44}); and
\item[H.]	age group 45-64 (\code{Age45_64}).
\end{itemize}

Thus the eight independent variables reflect measures of migration, public housing, high social class, unemployment, educational attainment, and three adult age groups.

The \code{EWHP} data (with $n=519$) is a house price data set for England and Wales, this time sampled from 1999, but again provided by the Nationwide Building Society and combined with various hedonic contextual variables.  Here for a regression fit, the dependent variable is \code{PurPrice} (what the house sold for) and the nine independent variables are: \code{BldIntWr}, \code{BldPostW}, \code{Bld60s}, \code{Bld70s}, \code{Bld80s}, \code{TypDetch}, \code{TypSemiD}, \code{TypFlat} and \code{FlrArea}.  All independent variables are indicator variables (1 or 0) except for \code{FlrArea}.  Section~\ref{sec.gwpred} uses this data when demonstrating GW regression as a spatial predictor; where \code{PurPrice} is considered as a function of \code{FlrArea} (house floor area), only.

%
%

\section[Distance matrix, kernel and bandwidth]{Distance matrix, kernel and bandwidth}\label{sec.distbw}
A fundamental element in GW modelling is the spatial weighting function \citep{FoBrCh:02a} that quantifies (or sets) the spatial relationship or spatial dependency between the observed variables.
Here $W(u_i,v_i)$ is a $n \times n$ (with $n$ the number of observations) diagonal matrix denoting the geographical weighting of each observation point for model calibration point $i$ at location $(u_i,v_i)$. 
We have a different diagonal matrix for each model calibration point. There are three key elements in building this weighting matrix: (i) the type of distance, (ii) the kernel function and (iii) its bandwidth.

\subsection[Selecting the distance function]{Selecting the distance function}\label{gw.dist}
Distance can be calculated in various ways and does not have to be Euclidean. An important family of distance metrics are Minkowski distances. This family includes the usual Euclidean distance having $p=2$ and the Manhattan distance when $p=1$ (where $p$ is the power of the Minkowski distance). Another useful metric is the great circle distance, which finds the shortest distance between two points taking into consideration the natural curvature of the Earth. All such metrics are possible in \pkg{GWmodel}.   

\subsection[Kernel functions and bandwidth]{Kernel functions and bandwidth}\label{gw.kbw}
\begin{figure}[!h]
    \centering
    \includegraphics[scale=.5]{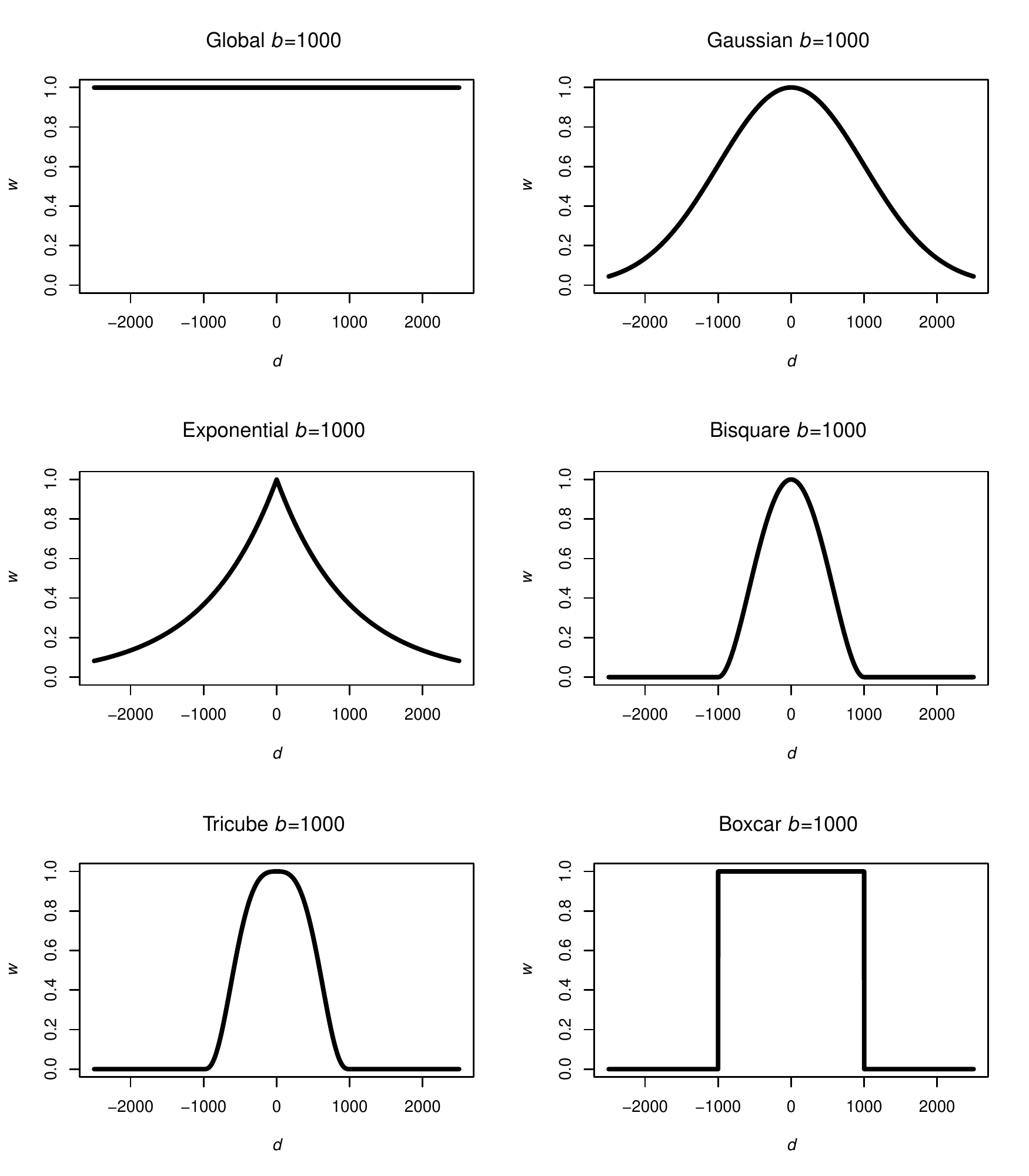} 
    \caption{Plot of the six kernel functions, with the bandwidth $b=1000$, and where $w$ is the weight, and $d$ is the distance between two observations}
    \label{plot.kernel}
\end{figure}

A set of commonly used kernel functions are shown in Table~\ref{tab.kernel} and Figure~\ref{plot.kernel}; all of which are available in \pkg{GWmodel}.
The `Global Model' kernel, that gives a unit weight to each observation, is included in order to show that a global model is a special case of its GW model.
\begin{table}[ht]
\centering
\begin{tabular}{lc}
  \hline
    \cr
  Global Model &
  $w_{ij} = 1$\\
  \cr
Gaussian &
$w_{ij} = \exp\left(-\frac{1}{2}\left(\frac{d_{ij}}{b}\right)^2\right)$\\
\cr
Exponential &
$w_{ij} = \exp\left(-\frac{|d_{ij}|}{b}\right)$
\\
\cr
Box-car & 
$w_{ij} = \left\{ 
  \begin{array}{c l}
   1 & \quad \text{if } |d_{ij}| < b,\\
   0 & \quad \text{otherwise}
  \end{array} \right.
$\\
\cr
Bi-square &
$w_{ij} = \left\{ 
  \begin{array}{c l}
   (1-(d_{ij}/b)^2)^2 & \quad \text{if } |d_{ij}| < b,\\
    0 & \quad \text{otherwise}
  \end{array} \right.
$
\\
\cr
Tri-cube &
$w_{ij} = \left\{ 
  \begin{array}{c l}
   (1-(|d_{ij}|/b)^3)^3 & \quad \text{if } |d_{ij}| < b,\\
    0 & \quad \text{otherwise}
  \end{array} \right.
$
\\
\cr   \hline
\end{tabular}
\caption{Six kernel functions; $w_{ij}$ is the $j$-th element of the diagonal of the matrix of geographical weights $W(u_i,v_i)$, and $d_{ij}$ is the distance between observations $i$ and $j$, and $b$ is the bandwidth.}
\label{tab.kernel}
\end{table}

The Gaussian and exponential kernels are continuous functions of the distance between two observation points (or an observation and calibration point).
The weights will be a maximum (equal to 1) for an observation at a GW model calibration point, and will decrease according to a Gaussian or exponential curve as the distance between observation/calibration points increases.

The box-car kernel is a simple discontinuous function that excludes observations that are further than some distance $b$ from the GW model calibration point.
This is equivalent to setting their weights to zero at such distances.
This kernel allows for efficient computation, since only a subset of the observation points need to be included in fitting the local model at each GW model calibration point.
This can be particularly useful when handling large data sets.

The bi-square and tri-cube kernels are similarly discontinuous, giving null weights to observations with a distance greater than $b$.
However unlike a box-car kernel, they provide weights that decrease as the distance between observation/calibration points increase, up until the distance $b$. Thus these are both distance-decay weighting kernels, as are Gaussian and exponential kernels.

The key controlling parameter in all kernel functions is the bandwidth $b$.
For the discontinuous functions, bandwidths can be specified either as a fixed distance or as a fixed number of local data (i.e., an adaptive distance).
For the continuous functions, bandwidths can be specified either as a fixed distance or as a `fixed quantity that reflects local sample size'  (i.e., still an `adaptive' distance, but the actual local sample size will be the sample size, as functions are continuous). In practise a fixed bandwidth suits fairly regular sample configurations whilst an adaptive bandwidth suits highly irregular sample configurations. Adaptive bandwidths ensure sufficient (and constant) local information for each local calibration of a given GW model. Bandwidths for GW models can be user-specified or found via some automated (e.g., cross-validation) procedure provided some objective function exists.
Specific functions (\code{bw.gwr}, \code{bw.gwr.lcr}, \code{bw.ggwr}, \code{bw.gwpca}, \code{bw.gwda}) can be used to find such optimal bandwidths, depending on the chosen GW model.\\ 

\subsection[Example]{Example}

As an example, we find the distance matrix for the house price data for England and Wales (\code{EWHP}), as described in Section~\ref{sec.dataset}. Here, the distance matrix can be calculated: (a) within a function of a specific GW model or (b) outside of the function and saved using the function \code{gw.dist}. This flexibilty is particularly useful for saving computation time when fitting several different GW models. Observe that we specify the Euclidean distance metric for this data. Other distance metrics could have been specified by: (1) modifying the parameter \code{p}, the power of the  Minkowsky distance or (2) setting \code{longlat=TRUE} for the great circle distance. The output of the function \code{gw.dist} is a matrix containing in each row the value of the diagonal of the distance matrix for each observation.

\begin{CodeChunk}
\begin{CodeInput}
R> library("GWmodel")
R> data("EWHP")
R> houses.spdf <- SpatialPointsDataFrame(ewhp[, 1:2], ewhp)
R> houses.spdf[1:6,]
\end{CodeInput}
\begin{CodeOutput}
  Easting Northing PurPrice BldIntWr BldPostW Bld60s Bld70s Bld80s TypDetch
1  599500   142200    65000        0        0      0      0      1        0
2  575400   167200    45000        0        0      0      0      0        0
3  530300   177300    50000        1        0      0      0      0        0
4  524100   170300   105000        0        0      0      0      0        0
5  426900   514600   175000        0        0      0      0      1        1
6  508000   190400   250000        0        1      0      0      0        1
  TypSemiD TypFlat   FlrArea
1        1       0  78.94786
2        0       1  94.36591
3        0       0  41.33153
4        0       0  92.87983
5        0       0 200.52756
6        0       0 148.60773
\end{CodeOutput}
\end{CodeChunk}
\begin{CodeChunk}
\begin{CodeInput}
R> DM <- gw.dist(dp.locat = coordinates(houses.spdf))
R> DM[1:7,1:7]
\end{CodeInput}
\begin{CodeOutput}
          [,1]      [,2]       [,3]       [,4]     [,5]      [,6]     [,7]
[1,]      0.00  34724.78  77592.848  80465.956 410454.0 103419.00 236725.0
[2,]  34724.78      0.00  46217.096  51393.579 377808.2  71281.13 202563.8
[3,]  77592.85  46217.10      0.000   9350.936 352792.9  25863.10 160741.1
[4,]  80465.96  51393.58   9350.936      0.000 357757.4  25753.06 160945.0
[5,] 410454.04 377808.17 352792.928 357757.362      0.0 334189.84 232275.4
[6,] 103419.00  71281.13  25863.101  25753.058 334189.8      0.00 135411.2
[7,] 236725.01 202563.77 160741.096 160945.022 232275.4 135411.23      0.0 
\end{CodeOutput}
\end{CodeChunk}

\section[GW summary statistics]{GW summary statistics} \label{sec.gwss}
This section presents the simplest form of GW modelling with GW summary statistics \citep{BrFoCh:02a,FoBrCh:02a}.  Here, we describe how to calculate GW means, GW standard deviations and GW measures of skew; which constitute a set of basic GW summary statistics.  To mitigate against any adverse effect of outliers on these local statistics, a set of robust alternatives are also described in GW medians, GW inter-quartile ranges and GW quantile imbalances. In addition, to such local \emph{univariate} summary statistics, GW correlations are described in basic and robust forms (Pearson's and Spearman's, respectively); providing a set of local \emph{bivariate} summary statistics.

Although fairly simple to calculate and map, GW summary statistics are considered a vital pre-cursor to an application of any subsequent GW model, such as a GW PCA (Section~\ref{sec.gwpca}) or GW regression (Sections~\ref{sec.gwr} to \ref{sec.gwpred}).  For example, GW standard deviations (or GW inter-quartile ranges) will highlight areas of high variability for a given variable, areas where a subsequent application of a GW PCA or a GW regression may warrant close scrutiny.  Basic and robust GW correlations provide a preliminary assessment of relationship non-stationarity between the dependent and an independent variable of a GW regression (Section~\ref{sec.gwr}).  GW correlations also provide an assessment of local collinearity between two independent variables of a GW regression; which could then lead to the application of a locally compensated model (Section~\ref{sec.multicoll}).

\subsection[Basic GW summary statistics]{Basic GW summary statistics}

For attributes $z$ and $y$ at any location $i$  where $w_{ij}$  accords to some kernel function of Section~\ref{sec.distbw}, definitions for a GW mean, a GW standard deviation, a GW measure of skew and a GW Pearson's correlation coefficient are respectively:

\begin{equation*}
m(z_i)=\frac{\sum_{j=1}^n w_{ij}z_j}{\sum_{j=1}^n w_{ij}}
\end{equation*}

\begin{equation*}
s(z_i)=\sqrt{\frac{\sum_{j=1}^n w_{ij}\left(z_j-m(z_i)\right)^2}{\sum_{j=1}^n w_{ij}}}
\end{equation*}

\begin{equation*}
b(z_i)=\frac{\left[\sqrt[3]{\frac{\sum_{j=1}^n w_{ij}\left(z_j-m(z_i)\right)^3}{\sum_{j=1}^n w_{ij}}}\right]}{s(z_i)}
\end{equation*}

and 

\begin{equation}\label{eq.sp}
\rho(z_i,y_i)=\frac{c(z_i,y_i)}{s(z_i)s(y_i)}
\end{equation} 

with the GW covariance:

\begin{equation*}
c(z_i,y_i)=\frac{\sum_{j=1}^n w_{ij}\left[\left(z_j-m(z_i) \right) \left(y_j-m(y_i) \right) \right]}{\sum_{j=1}^n w_{ij}}
\end{equation*}

\subsection[Robust GW summary statistics]{Robust GW summary statistics}

Definitions for a GW median, a GW inter-quartile range and a GW quantile imbalance, all require the calculation of GW quantiles at any location $i$; the calculation of which are detailed in \cite{BrFoCh:02a}.  Thus if we calculate GW quartiles, the GW median is the second GW quartile; and the GW inter-quartile range is the third minus the first GW quartile.  The GW quantile imbalance measures the symmetry of the middle part of the local distribution and is based on the position of the GW median relative to the first and third GW quartiles.  It ranges from -1 (when the median is very close to the first GW quartile) to 1 (when the median is very close to the third GW quartile), and is zero if the median bisects the first and third GW quartiles.  To find a GW Spearman's correlation coefficient, the local data for $z$ and for $y$, each need to be ranked using the same approach as that used to calculate the GW quantiles.  The locally ranked variables are then simply fed into Equation~\ref{eq.sp}.

\subsection[Example]{Example}
For demonstration of basic and robust GW summary statistics, we use the Dublin voter turnout data.  Here we investigate the local variability in voter turnout (\code{GenEl2004}), which is the dependent variable in the regressions of Sections~\ref{sec.gwr} and \ref{sec.multicoll}.  We also investigate the local relationships between: (i) turnout and \code{LARent} and (ii) \code{LARent} and \code{Unempl} (i.e., two independent variables in the regressions of Sections~\ref{sec.gwr} and \ref{sec.multicoll}).

For any GW model calibration, it is prudent to experiment with different kernel functions.  For our chosen GW summary statistics, we specify box-car and bi-square kernels; where the former relates to an un-weighted moving window, whilst the latter relates to a weighted one (from Section~\ref{sec.distbw}).  GW models using box-car kernels are useful in that the identification of outlying relationships or structures are more likely \citep{llsh06,HarBr10}.  Such calibrations more easily relate to the global model form (see Section~\ref{sec.multicoll}) and in turn, tend to provide an intuitive understanding of the degree of heterogeneity in the process.  Observe that it is always possible that the spatial process is essentially homogeneous, and in such cases, the output of a GW model can confirm this.

The spatial arrangement of the EDs in Greater Dublin is not a tessellation of equally sized zones, so it makes sense to specify an adaptive kernel bandwidth.  For example, if we specify a bandwidth of $N=100$, the box-car and bi-square kernels will change in radius but will always include the closest 100 EDs for each local summary statistic.  We also user-specify the bandwidths.  Note that bandwidths for GW means or medians can be found optimally via cross-validation, as an objective function exists.  That is, 'leave-one-out' predictions can be found and compared to the actual data, across a range of bandwidths.  The optimal bandwidth is that which provides the most accurate predictions. However, such bandwidth selection functions are not yet incorporated in \pkg{GWmodel}. For all other GW summary statistics, bandwidths can only be user-specified, as no objective function exists for them.

Commands to conduct our local analysis are as follows, where we use the function \code{gwss} with two different specifications to find our GW summary statistics.  We specify box-car and bi-square kernels, each with an adaptive bandwidth of $N=48$ (approximately 15\% of the data).  To find robust GW summary statistics based on quantiles, the \code{gwss} function is specified with \code{quantiles = TRUE} (observe that we do not need to do this for our robust GW correlations).

\begin{CodeChunk}
\begin{CodeInput}
R> data("DubVoter")

R> gw.ss.bx <- gwss(Dub.voter, vars  =  c("GenEl2004", "LARent", "Unempl"),
+ kernel = "boxcar", adaptive = TRUE, bw = 48, quantile = TRUE)
R> gw.ss.bs <- gwss(Dub.voter,vars = c("GenEl2004", "LARent", "Unempl"),
+ kernel = "bisquare", adaptive = TRUE, bw = 48)
\end{CodeInput}
\end{CodeChunk}

\begin{figure}
\centering
\begin{subfigure}{.5\textwidth}
  \centering
  \includegraphics[width=1\linewidth]{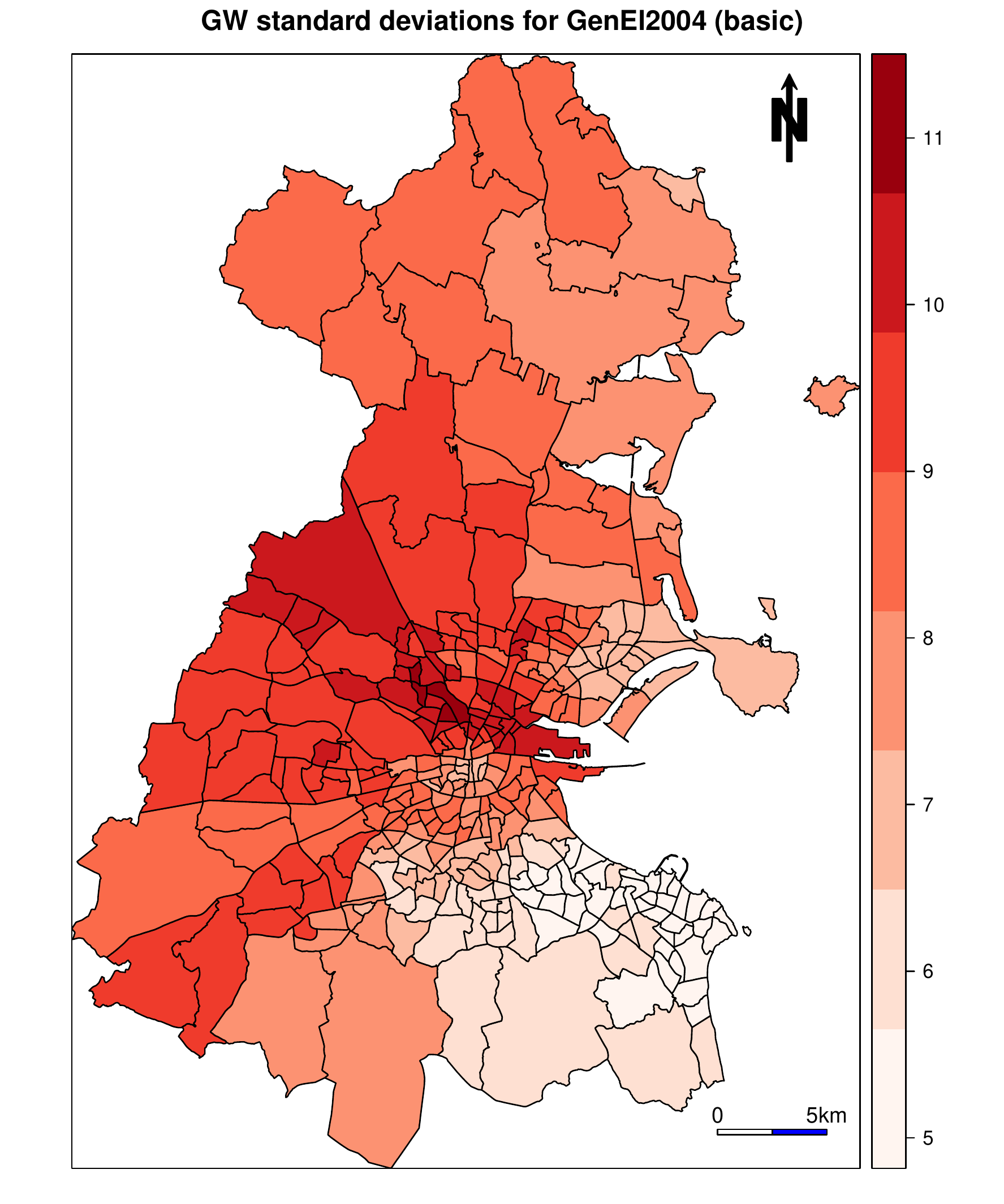}
  \caption{}
  \label{fig:gwss_2a}
\end{subfigure}%
\begin{subfigure}{.5\textwidth}
  \centering
  \includegraphics[width=1\linewidth]{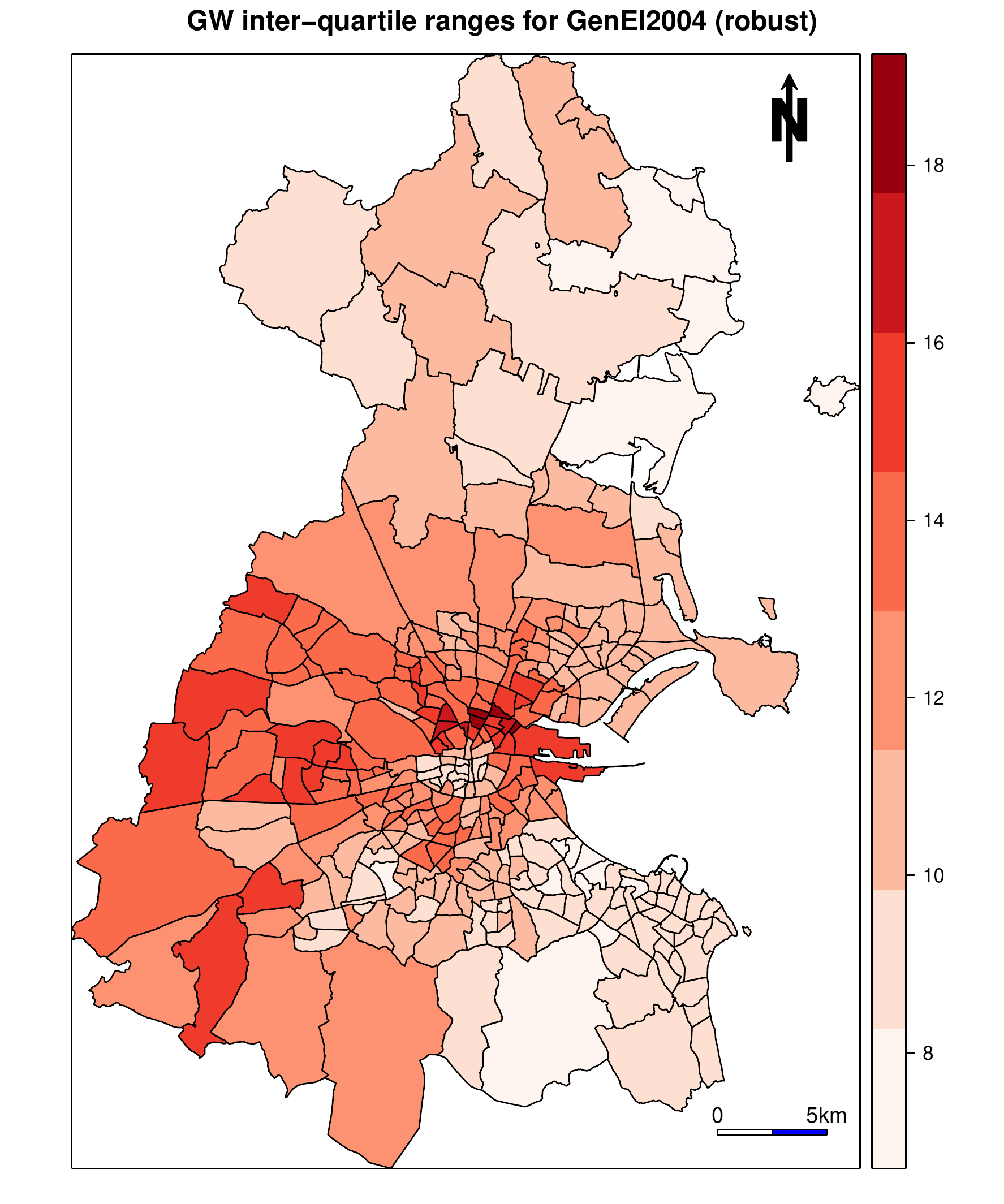}
  \caption{}
  \label{fig:gwss_2b}
\end{subfigure}
\caption{(a) Basic and (b) robust GW measures of variability for \code{GenEl2004} (turnout).}
\label{fig:gwss2}
\end{figure}

\begin{figure}
\centering
\begin{subfigure}{.5\textwidth}
  \centering
  \includegraphics[width=1\linewidth]{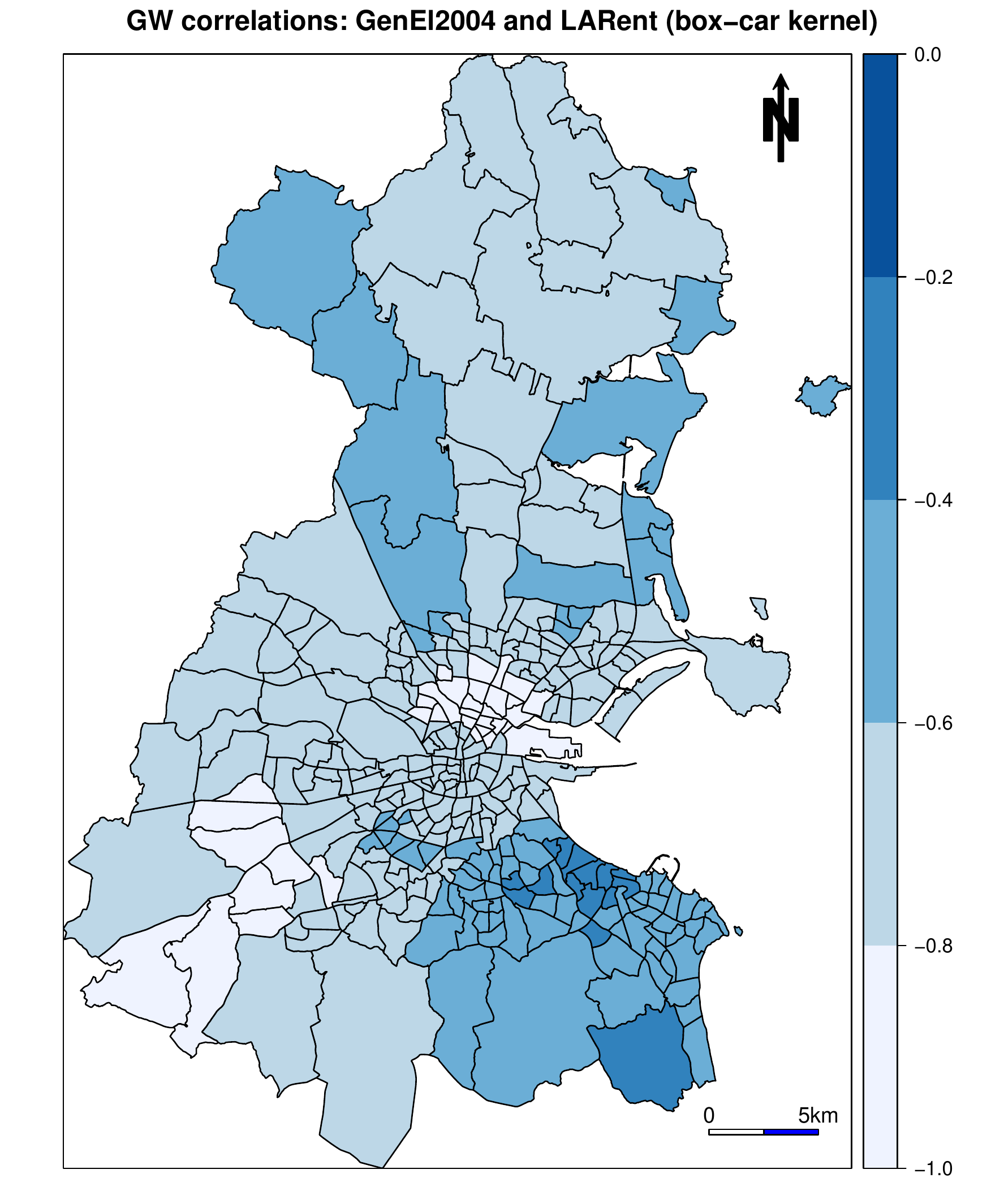}
  \caption{}
  \label{fig:gwss_3a}
\end{subfigure}%
\begin{subfigure}{.5\textwidth}
  \centering
  \includegraphics[width=1\linewidth]{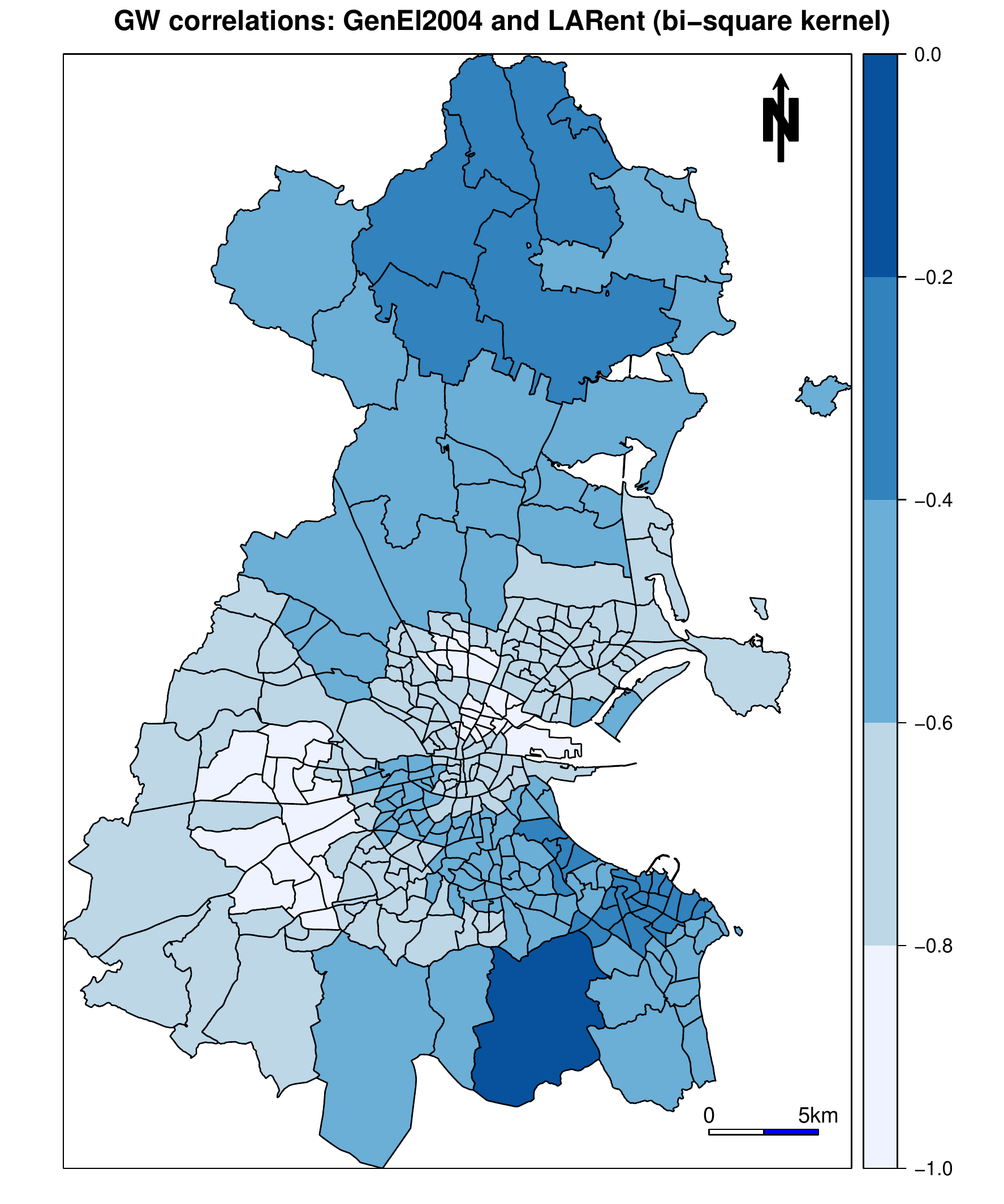}
  \caption{}
  \label{fig:gwss_3b}
\end{subfigure}
\caption{(a) Box-car and (b) bi-square specified GW correlations for \code{GenEl2004} and \code{LARent}.}
\label{fig:gwss3}
\end{figure}

\begin{figure}
\centering
\begin{subfigure}{.5\textwidth}
  \centering
  \includegraphics[width=1\linewidth]{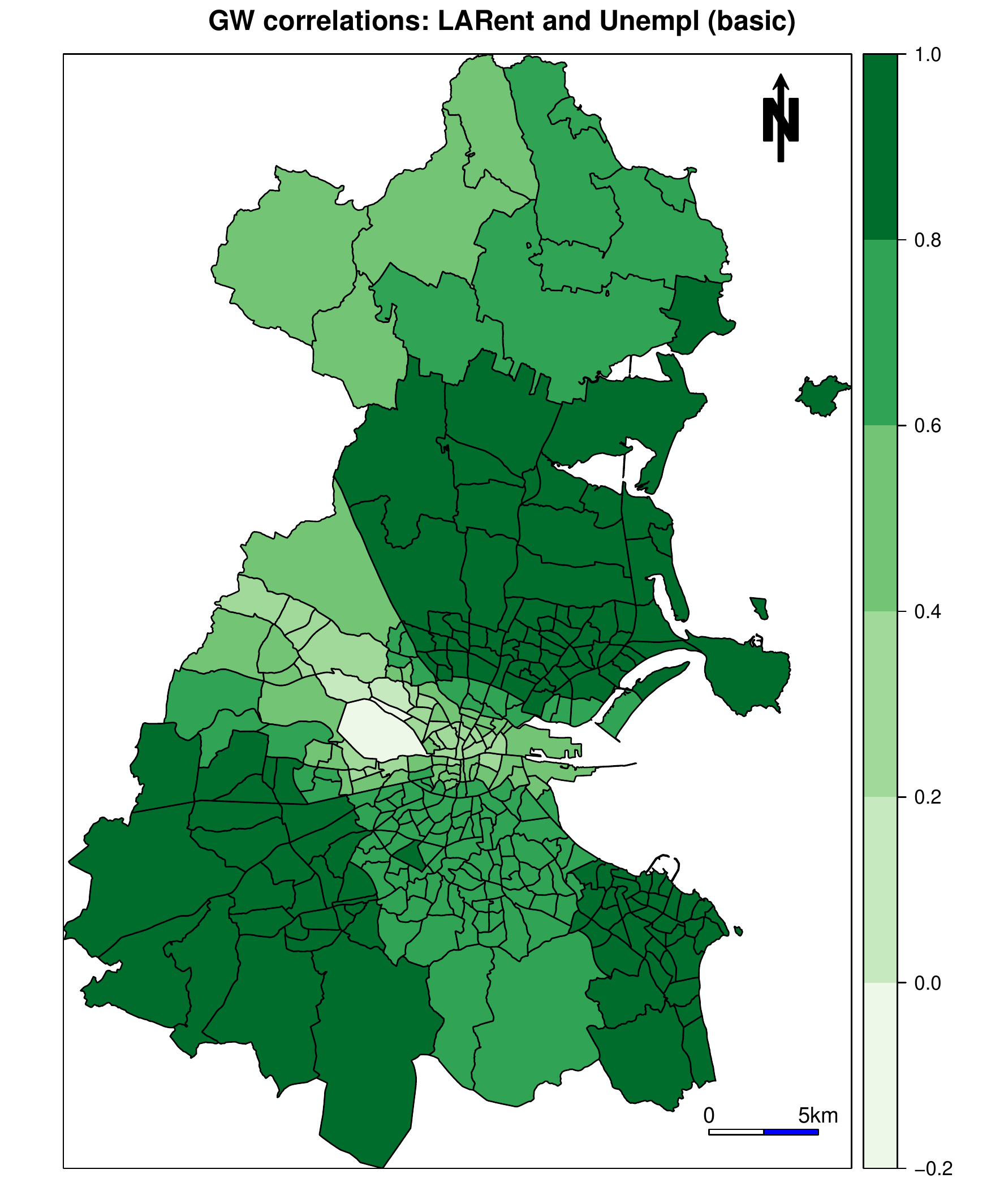}
  \caption{}
  \label{fig:gwss_4a}
\end{subfigure}%
\begin{subfigure}{.5\textwidth}
  \centering
  \includegraphics[width=1\linewidth]{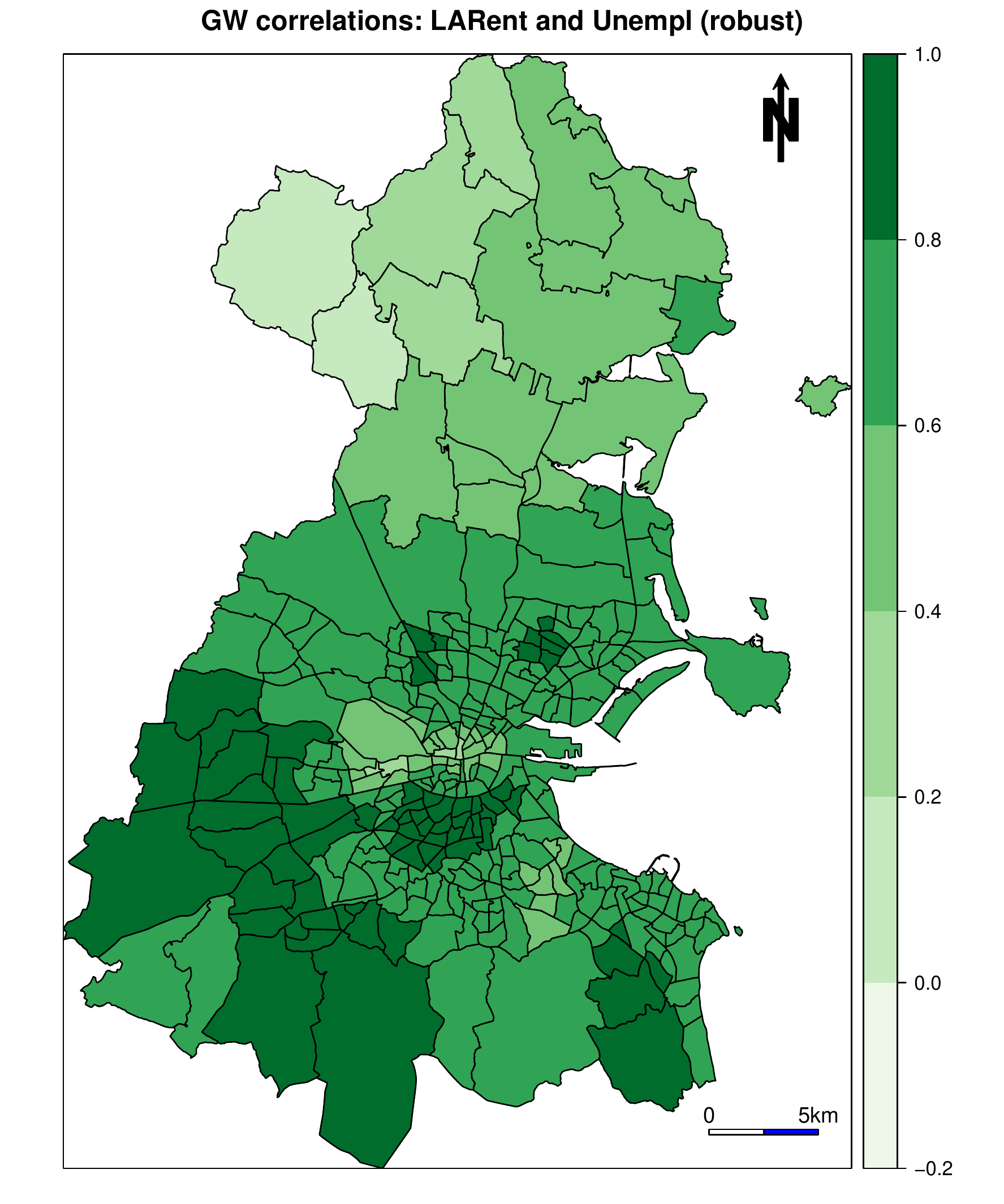}
  \caption{}
  \label{fig:gwss_4b}
\end{subfigure}
\caption{(a) Basic and (b) robust GW correlations for \code{LARent} and \code{Unempl}.}
\label{fig:gwss4}
\end{figure}

From these calibrations, we present three pairs of example visualisations: (a) basic and robust GW measures of variability for \code{GenEl2004} (each using a box-car kernel) in Figure~\ref{fig:gwss2}; (b) box-car and bi-square specified (basic) GW correlations for \code{GenEl2004} and \code{LARent} in Figure~\ref{fig:gwss3}; and (c) basic and robust GW correlations for \code{LARent} and \code{Unempl} (each using a bi-square kernel) in Figure~\ref{fig:gwss4}. Commands to conduct these visualisations are as follows:

\begin{CodeChunk}
\begin{CodeInput}
R> library("RColorBrewer")

R> map.na = list("SpatialPolygonsRescale", layout.north.arrow(),
+ offset = c(329000, 261500), scale = 4000, col = 1)
R> map.scale.1 = list("SpatialPolygonsRescale", layout.scale.bar(),
+ offset = c(326500, 217000), scale = 5000, col = 1, 
+ fill = c("transparent", "blue"))
R> map.scale.2  = list("sp.text", c(326500, 217900), "0", cex = 0.9, col = 1)
R> map.scale.3  = list("sp.text", c(331500, 217900), "5km", cex = 0.9, col = 1)
R> map.layout <- list(map.na, map.scale.1, map.scale.2, map.scale.3)

R> mypalette.1 <- brewer.pal(8, "Reds")
R> mypalette.2 <- brewer.pal(5, "Blues")
R> mypalette.3 <- brewer.pal(6, "Greens")

R> X11(width = 10, height = 12)
R> spplot(gw.ss.bx$SDF, "GenEl2004_LSD", key.space = "right",
+ col.regions = mypalette.1, cuts = 7,
+ main = "GW standard deviations for GenEl2004 (basic)",
+ sp.layout = map.layout)

R> X11(width = 10, height = 12)
R> spplot(gw.ss.bx$SDF, "GenEl2004_IQR", key.space = "right",
+ col.regions = mypalette.1, cuts = 7,
+ main = "GW inter-quartile ranges for GenEl2004 (robust)",
+ sp.layout = map.layout)

R> X11(width = 10, height = 12)
R> spplot(gw.ss.bx$SDF, "Corr_GenEl2004.LARent", key.space = "right",
+ col.regions = mypalette.2, at = c(-1, -0.8, -0.6, -0.4, -0.2, 0), 
+ main = "GW correlations: GenEl2004 and LARent (box-car kernel)",
+ sp.layout = map.layout)

R> X11(width = 10, height = 12)
R> spplot(gw.ss.bs$SDF, "Corr_GenEl2004.LARent", key.space = "right",
+ col.regions = mypalette.2, at = c(-1, -0.8, -0.6, -0.4, -0.2, 0),
+ main = "GW correlations: GenEl2004 and LARent (bi-square kernel)",
+ sp.layout = map.layout)

R> X11(width = 10, height = 12)
R> spplot(gw.ss.bs$SDF, "Corr_LARent.Unempl" ,key.space = "right",
+ col.regions = mypalette.3, at = c(-0.2, 0, 0.2, 0.4, 0.6, 0.8, 1),
+ main = "GW correlations: LARent and Unempl (basic)",
+ sp.layout = map.layout)

R> X11(width = 10, height = 12)
R> spplot(gw.ss.bs$SDF, "Spearman_rho_LARent.Unempl",key.space = "right",
+ col.regions = mypalette.3, at = c(-0.2, 0, 0.2, 0.4, 0.6, 0.8, 1),
+ main = "GW correlations: LARent and Unempl (robust)",
+ sp.layout = map.layout)
\end{CodeInput}
\end{CodeChunk}

From Figure~\ref{fig:gwss2}, we can see that turnout appears highly variable in areas of central and west Dublin.  From Figure~\ref{fig:gwss3}, the relationship between turnout and \code{LARent} appears non-stationary, where this relationship is strongest in areas of central and south-west Dublin.  Here turnout tends to be low while local authority renting tends to be high.  From Figure~\ref{fig:gwss4}, consistently strong positive correlations between \code{LARent} and \code{Unempl} are found in south-west Dublin.  This is precisely an area of Dublin where local collinearity in the GW regression of Section~\ref{sec.multicoll} is found to be strong and a cause for concern.

From these visualisations, it is clearly important to experiment with the calibration of a GW model, as subtle differences in our perception of the non-stationary effect can result by a simple altering of the specification.  Experimentation with different bandwidth sizes is also important, especially in cases when an optimal bandwidth cannot be specified. Observe that all GW models are primarily viewed as exploratory spatial data analysis (ESDA) tools and as such, experimentation is a vital aspect of this.

\section[GW principal components analysis]{GW principal components analysis}\label{sec.gwpca}
Principal components analysis (PCA) is a key method for the analysis of multivariate data \citep[see][]{jol02}.  A member of the unconstrained ordination family, it is commonly used to explain the covariance structure of a (high-dimensional) multivariate data set using only a few components (i.e., provide a low-dimensional alternative).  The components are linear combinations of the original variables and can potentially provide a better understanding of differing sources of variation and structure in the data.  These may be visualised and interpreted using associated graphics.
In geographical settings, standard PCA, in which the components do not depend on location, may be replaced with a GW PCA \citep{FoBrCh:02a,lloy:10a,gwpca11}, to account for spatial heterogeneity in the structure of the multivariate data.  In doing so, GW PCA can identify regions where assuming the same underlying structure in all locations is inappropriate or over-simplistic.  GW PCA can assess: (i) how (effective) data dimensionality varies spatially and (ii) how the original variables influence each spatially-varying component.  In part, GW PCA resembles the bivariate GW correlations of Section~\ref{sec.gwss} in a multivariate sense, as both are unlike the multivariate GW regressions of Sections~\ref{sec.gwr} to \ref{sec.gwpred}, since there is no distinction between dependent and independent variables.  Key challenges in GW PCA are: (a) finding the scale at which each localised PCA should operate and (b) visualising and interpreting the output that results from its application.  As with any GW model, GW PCA is constructed using weighted data that is controlled by the kernel function and its bandwidth (Section~\ref{sec.distbw}).

\subsection[GW PCA]{GW PCA}
More formally, for a vector of observed variables $x_i$ at spatial location $i$   with coordinates $(u,v)$, GW PCA involves regarding $x_i$ as conditional on $u$ and $v$, and making the mean vector $\mu$ and covariance matrix $\Sigma$, functions of $u$ and  $v$.  That is, $\mu(u,v)$ and $\Sigma(u,v)$ are the local mean vector and the local covariance matrix, respectively.  To find the local principal components, the decomposition of the local covariance matrix provides the local eigenvalues and local eigenvectors.  The product of the $i$-th row of the data matrix with the local eigenvectors for the $i$-th location provides the $i$-th row of local component scores.  The local covariance matrix is:
\begin{equation*}
\Sigma(u,v)=X^\top W(u,v)X
\end{equation*}
where $X$ is the data matrix (with $n$ rows for the observations and $m$ columns for the variables); and $W(u,v)$ is a diagonal matrix of geographic weights.  The local principal components at location $(u_i,v_i)$ can be written as:
\begin{equation*}
L(u_i,v_i)V(u_i,v_i)L(u_i,v_i)^\top=\Sigma(u_i,v_i)
\end{equation*}
where $L(u_i,v_i)$ is a matrix of local eigenvectors; $V(u_i,v_i)$ is a diagonal matrix of local eigenvalues; and $\Sigma(u_i,v_i)$ is the local covariance matrix.  Thus for a GW PCA with $m$ variables, there are $m$ components, $m$ eigenvalues, $m$ sets of component scores, and $m$ sets of component loadings at each observed location.  We can also obtain eigenvalues and their associated eigenvectors at un-observed locations, although as no data exists for these locations, we cannot obtain component scores.

\subsection[Robust GW PCA]{Robust GW PCA}
A robust GW PCA can also be specified, so as to reduce the effect of anomalous observations on its outputs.  Outliers can artificially increase local variability and mask key features in local data structures.  To provide a robust GW PCA, each local covariance matrix is estimated using the robust minimum covariance determinant (MCD) estimator \citep{rou85}.  The MCD estimator searches for a subset of $h$  data points that has the smallest determinant for their basic sample covariance matrix.  Crucial to the robustness and efficiency of this estimator is $h$, and we specify a default value of $h=0.75n$, following the recommendation of \cite{vafi09}.

\subsection[Example]{Example}
For applications of PCA and GW PCA, we again use the Dublin voter turnout data, this time focussing on the eight variables: \code{DiffAdd}, \code{LARent}, \code{SC1}, \code{Unempl}, \code{LowEduc}, \code{Age18_24}, \code{Age25_44} and \code{Age45_64} (i.e., the independent variables of the regression fits in Sections~\ref{sec.gwr} and \ref{sec.multicoll}).  Although measured on the same scale, the variables are not of a similar magnitude.  Thus, we standardise the data and specify our PCA with the covariance matrix.  The same (globally) standardised data is also used in our GW PCA calibrations, which are similarly specified with (local) covariance matrices.  The effect of this standardisation is to make each variable have equal importance in the subsequent analysis (at least for the PCA case)\footnote{The use of un-standardised data, or the use of locally-standardised data with GW PCA is a subject of current investigation.}.  The basic and robust PCA results are found using \code{scale}, \code{princomp} and \code{covMcd} functions, as follows:

\begin{CodeChunk}
\begin{CodeInput}
R> Data.scaled <- scale(as.matrix(Dub.voter@data[,4:11]))

R> pca.basic <- princomp(Data.scaled, cor = F)
R> (pca.basic$sdev^2 / sum(pca.basic$sdev^2))*100
\end{CodeInput}
\begin{CodeOutput}
   Comp.1    Comp.2    Comp.3    Comp.4    Comp.5    Comp.6    Comp.7   
36.084435 25.586984 11.919681 10.530373  6.890565  3.679812  3.111449  
   Comp.8
 2.196701
\end{CodeOutput}
\begin{CodeInput}
R> pca.basic$loadings
\end{CodeInput}
\begin{CodeOutput}
Loadings:
         Comp.1 Comp.2 Comp.3 Comp.4 Comp.5 Comp.6 Comp.7 Comp.8
DiffAdd   0.389 -0.444        -0.149  0.123  0.293  0.445  0.575
LARent    0.441  0.226  0.144  0.172  0.612  0.149 -0.539  0.132
SC1      -0.130 -0.576        -0.135  0.590 -0.343        -0.401
Unempl    0.361  0.462         0.189  0.197         0.670 -0.355
LowEduc   0.131  0.308 -0.362 -0.861                            
Age18_24  0.237         0.845 -0.359 -0.224               -0.200
Age25_44  0.436 -0.302 -0.317        -0.291  0.448 -0.177 -0.546
Age45_64 -0.493  0.118  0.179 -0.144  0.289  0.748  0.142 -0.164
\end{CodeOutput}
\begin{CodeInput}
R> R.COV <- covMcd(Data.scaled, cor = F, alpha = 0.75)
R> pca.robust <- princomp(Data.scaled, covmat = R.COV, cor = F)
R> pca.robust$sdev^2 / sum(pca.robust$sdev^2)
\end{CodeInput}
\begin{CodeOutput}
     Comp.1      Comp.2      Comp.3      Comp.4      Comp.5 
0.419129445 0.326148321 0.117146840 0.055922308 0.043299600 
     Comp.6      Comp.7      Comp.8 
0.017251964 0.014734597 0.006366926 
\end{CodeOutput}
\begin{CodeInput}
R> pca.robust$loadings
\end{CodeInput}
\begin{CodeOutput}
Loadings:
         Comp.1 Comp.2 Comp.3 Comp.4 Comp.5 Comp.6 Comp.7 Comp.8
DiffAdd   0.512        -0.180  0.284 -0.431  0.659              
LARent          -0.139         0.310  0.119               -0.932
SC1       0.559  0.591  0.121  0.368  0.284 -0.324              
Unempl   -0.188 -0.394         0.691        -0.201  0.442  0.307
LowEduc  -0.102 -0.186         0.359               -0.895  0.149
Age18_24               -0.937         0.330                     
Age25_44  0.480 -0.437        -0.211 -0.407 -0.598              
Age45_64 -0.380  0.497 -0.264  0.178 -0.665 -0.241
\end{CodeOutput}
\end{CodeChunk}
From the `percentage of total variance' (PTV) results, the first three components collectively account for 73.6\% and 86.2\% of the variation in the data, for the basic and robust PCA, respectively.  From the tables of loadings, component one would appear to represent older residents (\code{Age45_64}) in the basic PCA or represent affluent residents (\code{SC1}) in the robust PCA.  Component two, appears to represent affluent residents in both the basic and robust PCA.  These are whole-map statistics \citep{opchwycr87} and interpretations that represent a Dublin-wide average.  However, it is possible that they do not represent local social structure particularly reliably. If this is the case, an application of GW PCA may be useful, which will now be demonstrated.

Kernel bandwidths for GW PCA can be found automatically using a cross-validation approach, similar in nature to that used in GW regression (Section~\ref{sec.gwr}).  Details of this automated procedure are described in \cite{gwpca11}, where, a `leave-one-out' cross-validation (CV) score is computed for all possible bandwidths and an optimal bandwidth relates to the smallest CV score found.  With this procedure, it is currently necessary to decide \emph{a priori} upon the number of components to retain ($k$, say), and a different optimal bandwidth results for each $k$.  The procedure does not yield an optimal bandwidth if all components are retained (i.e., $m=k$); in this case, the bandwidth must be user-specified.  Thus for our analysis, an optimal adaptive bandwidth is found using a bi-square kernel, for both a basic and a robust GW PCA. Here, $k=3$ is chosen on an \emph{a priori} basis.  With \pkg{GWmodel}, the \code{bw.gwpca} function is used in the following set of commands, where the standardised data is converted to a spatial form via the \code{SpatialPointsDataFrame} function.

\begin{CodeChunk}
\begin{CodeInput}
R> Coords <- as.matrix(cbind(Dub.voter$X, Dub.voter$Y))
R> Data.scaled.spdf <- 
+ SpatialPointsDataFrame(Coords,as.data.frame(Data.scaled))

R> bw.gwpca.basic <- bw.gwpca(Data.scaled.spdf,
+ vars = colnames(Data.scaled.spdf@data), k = 3, robust = FALSE,
+ adaptive = TRUE)
R> bw.gwpca.basic
\end{CodeInput}
\begin{CodeOutput}
[1] 131
\end{CodeOutput}
\begin{CodeInput}
R> bw.gwpca.robust <- bw.gwpca(Data.scaled.spdf,
+ vars=colnames(Data.scaled.spdf@data), k = 3, robust = TRUE, 
+ adaptive = TRUE)
R> bw.gwpca.robust
\end{CodeInput}
\begin{CodeOutput}
[1] 130
\end{CodeOutput}
\end{CodeChunk}

Inspecting the values of \code{bw.gwpca.basic} and \code{bw.gwpca.robust} show that (very similar) optimal bandwidths of $N=131$ and $N=130$ will be used to calibrate the respective basic and robust GW PCA fits.  Observe that we now specify all $k=8$ components, but will focus our investigations on only the first three components.  This specification ensures that the variation locally accounted for by each component, is estimated correctly.  The two GW PCA fits are found using the \code{gwpca} function as follows:

\begin{CodeChunk}
\begin{CodeInput}
R> gwpca.basic <- gwpca(Data.scaled.spdf,
+ vars = colnames(Data.scaled.spdf@data), bw = bw.gwpca.basic, k = 8, 
+ robust = FALSE, adaptive = TRUE)

R> gwpca.robust <- gwpca(Data.scaled.spdf, 
+ vars = colnames(Data.scaled.spdf@data), bw = bw.gwpca.robust, k = 8, 
+ robust = TRUE, adaptive = TRUE)
\end{CodeInput}
\end{CodeChunk}
The GW PCA outputs\footnote{For a more objective and direct comparison of the basic and robust fits, the use of the same bandwidth (say that found optimally for the basic fit) may be preferable.} may now be now visualised and interpreted, focusing on: (1) how data dimensionality varies spatially and (2) how the original variables influence the components.  For the former, the spatial distribution of local PTV for say, the first three components can be mapped.  Commands to conduct this mapping for basic and robust GW PCA outputs are as follows, where the \code{prop.var} function is used to find the PTV data, which is then added to the Dub.voter spatial data frame, so that it can be easily mapped using the \code{spplot} function.
\begin{CodeChunk}
\begin{CodeInput}
R> prop.var <- function(gwpca.obj, n.components) {
+ return((rowSums(gwpca.obj$var[, 1:n.components])/
+ rowSums(gwpca.obj$var))*100)
+ }

R> var.gwpca.basic <- prop.var(gwpca.basic, 3) 
R> var.gwpca.robust <- prop.var(gwpca.robust, 3)

R> Dub.voter$var.gwpca.basic <- var.gwpca.basic
R> Dub.voter$var.gwpca.robust <- var.gwpca.robust

R> mypalette.4 <-brewer.pal(8, "YlGnBu")

 
R> X11(width = 10,height = 12)
R> spplot(Dub.voter, "var.gwpca.basic", key.space = "right",
+ col.regions = mypalette.4, cuts = 7, 
+ main = "PTV for local components 1 to 3 (basic GW PCA)", 
+ sp.layout = map.layout)

R> X11(width = 10,height = 12)
R> spplot(Dub.voter, "var.gwpca.robust", key.space = "right",
+ col.regions = mypalette.4, cuts = 7,
+ main = "PTV for local components 1 to 3 (robust GW PCA)", 
+ sp.layout = map.layout)
\end{CodeInput}
\end{CodeChunk}

\begin{figure}
\centering
\begin{subfigure}{.5\textwidth}
  \centering
  \includegraphics[width=1\linewidth]{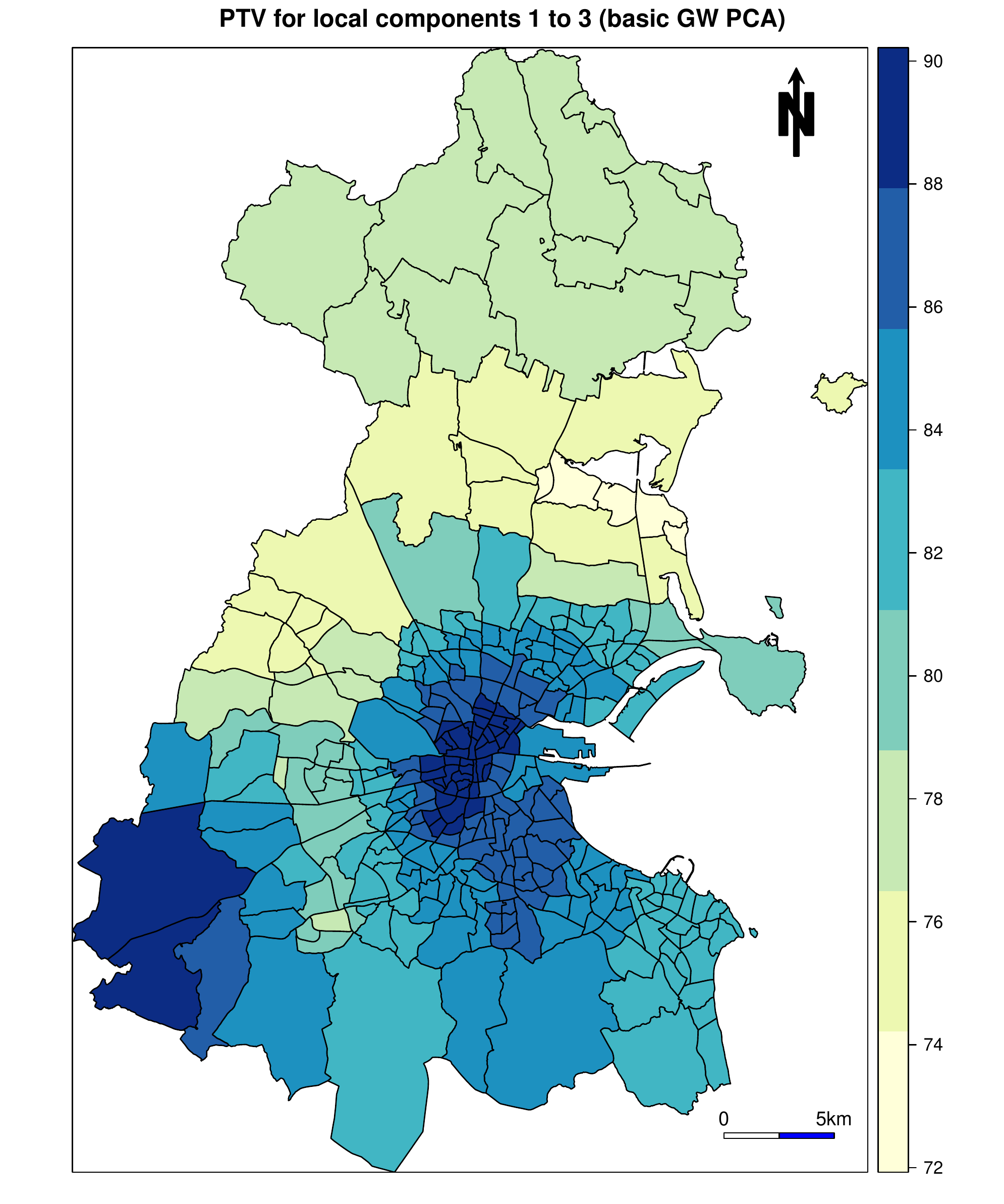}
  \caption{}
  \label{fig:gwpca_5a}
\end{subfigure}%
\begin{subfigure}{.5\textwidth}
  \centering
  \includegraphics[width=1\linewidth]{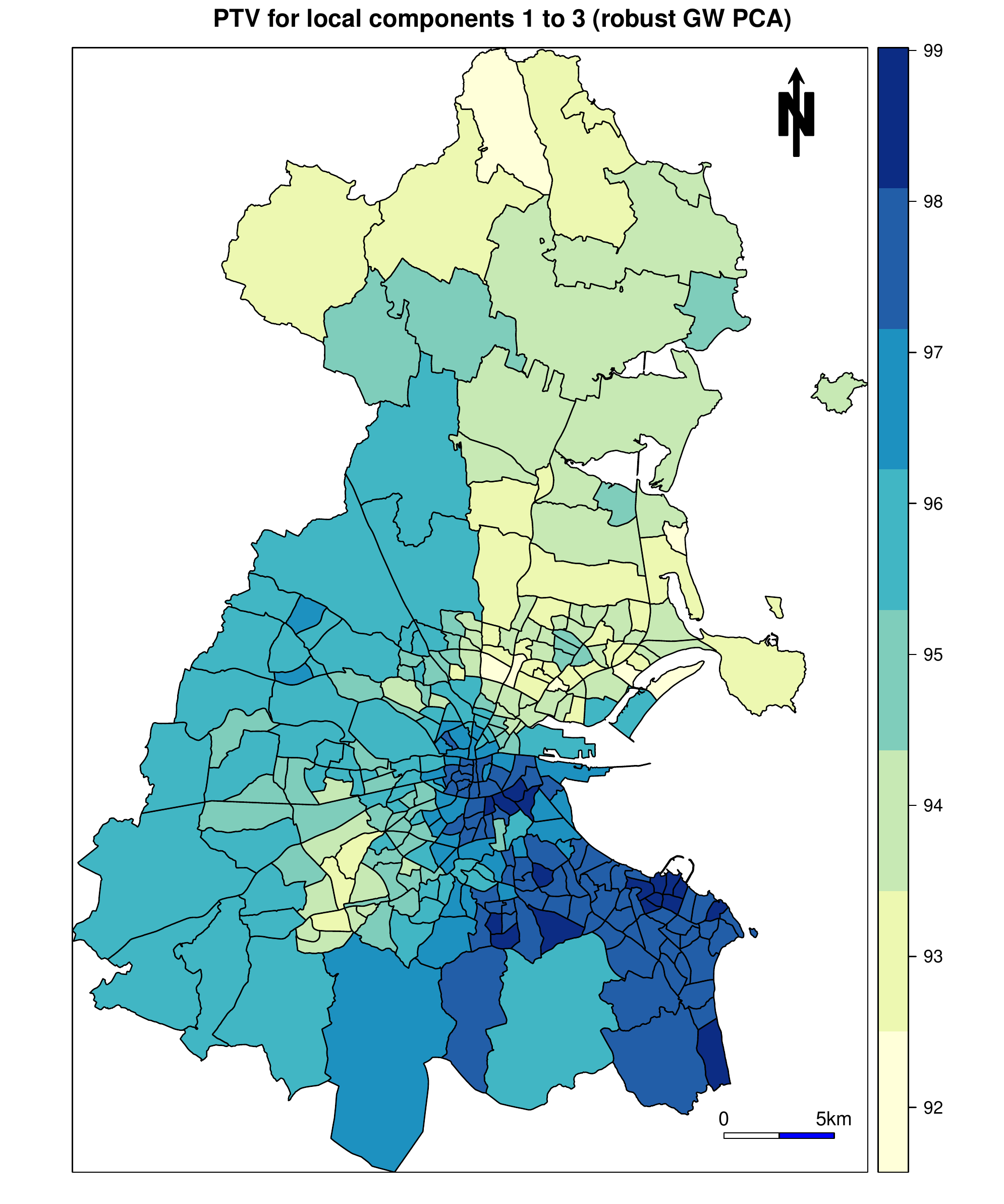}
  \caption{}
  \label{fig:gwpca_5b}
\end{subfigure}
\caption{(a) Basic and (b) robust PTV data for the first three local components.}
\label{fig:gwpca5}
\end{figure}

Figure~\ref{fig:gwpca5} presents the local PTV maps for the two GW PCA fits.  There is clear geographical variation in the PTV data and a higher PTV is generally accounted for in the local case, than in the global case.  The spatial patterns in both maps are broadly similar, with higher percentages located in the south, whilst lower percentages are located in the north.  As would be expected, the robust PTV data is consistently higher than the basic PTV data.  Variation in the basic PTV data is also greater than that found in the robust PTV data.  Large (relative) differences between the basic and robust PTV outputs (e.g., in south-west Dublin) can be taken to indicate the existence of global or possibly, local multivariate outliers.

We can next visualise how each of the eight variables locally influence a given component, by mapping the 'winning variable' with the highest absolute loading.  For brevity, we present such maps for the first component, only.  Commands to conduct this mapping for basic and robust GW PCA outputs are as follows:

\begin{CodeChunk}
\begin{CodeInput}
R> loadings.pc1.basic <- gwpca.basic$loadings[,,1]
R> win.item.basic = max.col(abs(loadings.pc1.basic))
 
R> loadings.pc1.robust <- gwpca.robust$loadings[,,1]
R> win.item.robust = max.col(abs(loadings.pc1.robust))

R> Dub.voter$win.item.basic <- win.item.basic
R> Dub.voter$win.item.robust <- win.item.robust
 
R> mypalette.5 <- c("lightpink", "blue", "grey", "purple",
+ "orange", "green", "brown", "yellow")
 
R> X11(width = 10,height = 12)
R> spplot(Dub.voter, "win.item.basic", key.space = "right",
+ col.regions = mypalette.5, at = c(1, 2, 3, 4, 5, 6, 7, 8, 9),
+ main = "Winning variable: highest abs. loading on local Comp.1 (basic)",
+ colorkey = F, sp.layout = map.layout)
 
R> X11(width = 10,height = 12)
R> spplot(Dub.voter, "win.item.robust", key.space = "right",
+ col.regions = mypalette.5, at = c(1, 2, 3, 4, 5, 6, 7, 8, 9),
+ main = "Winning variable: highest abs. loading on local Comp.1 (robust)",
+ colorkey = F, sp.layout = map.layout)
\end{CodeInput}
\end{CodeChunk}

\begin{figure}[!ht]
\centering
\begin{subfigure}{.5\textwidth}
  \centering
  \includegraphics[width=1\linewidth]{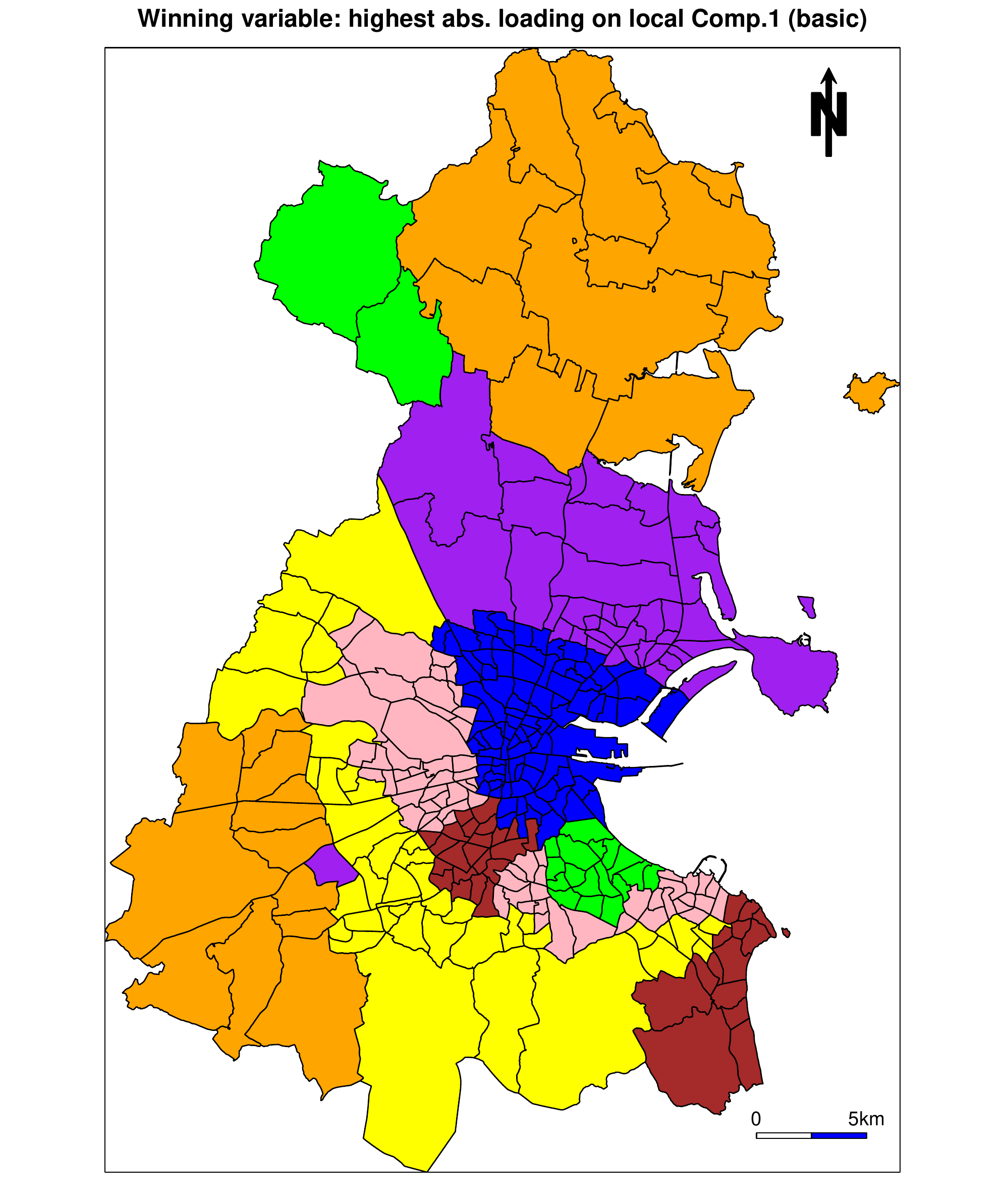}
  \caption{}
  \label{fig:gwpca_6a}
\end{subfigure}%
\begin{subfigure}{.5\textwidth}
  \centering
  \includegraphics[width=1\linewidth]{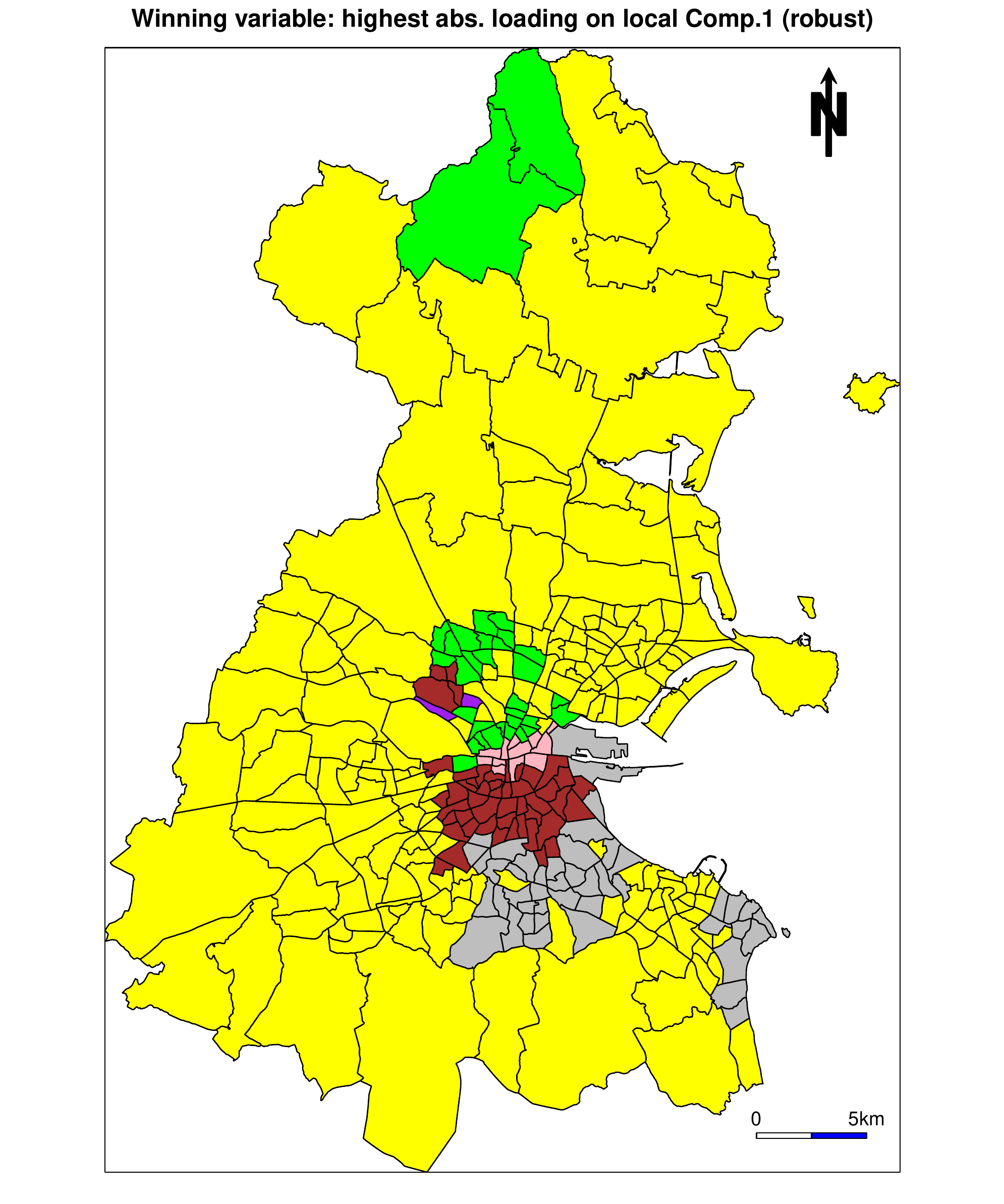}
  \caption{}
  \label{fig:gwpca_6b}
\end{subfigure}
\caption{(a) Basic and (b) robust GW PCA results for the winning variable on the first component. Map legends are: \code{DiffAdd} - light pink; \code{LARent} - blue; \code{SC1} - grey; \code{Unempl} - purple; \code{LowEduc} - orange; \texttt{Age18\_24} - green; \texttt{Age25\_44} - brown; and \texttt{Age45\_64} - yellow.}
\label{fig:gwpca6}
\end{figure}

Figure~\ref{fig:gwpca6} presents the `winning variable' maps for the two GW PCA fits, where we can observe clear geographical variation in the influence of each variable on the first component.  For basic GW PCA, low educational attainment (\code{Low_Educ}) dominates in the northern and south-western EDs, whilst public housing (\code{LARent}) dominates in the EDs of central Dublin.  The corresponding PCA `winning variable' is \code{Age45_64}, which is clearly not dominant throughout Dublin.  Variation in the results from basic GW PCA is much greater than that found with robust GW PCA (reflecting analogous results to that found with the PTV data).  For robust GW PCA, \code{Age45_64} does in fact dominate in most areas, thus reflecting a closer correspondence to the global case - but interestingly only the basic fit, and not the robust fit.

\section[GW regression]{GW regression}\label{sec.gwr}
\subsection[Basic GW regression]{Basic GW regression}

The most popular GW model is GW regression \citep{BrFoCh:96a,BrFoCh:98}, where spatially-varying relationships are explored between the dependent and independent variables.  Exploration commonly consists of mapping the resultant local regression coefficient estimates and associated (pseudo) \emph{t}-values to determine evidence of non-stationarity.  The basic form of the GW regression model is:
\begin{equation*}
y_i=\beta_{i0}+\sum_{k=1}^{m}\beta_{ik}x_{ik}+\epsilon_i
\end{equation*}
where $y_i$ is the dependent variable at location $i$; $x_{ik}$ is the value of the $k$th independent variable at location $i$; $m$ is the number of independent variables; $\beta_{i0}$ is the intercept parameter at location $i$; $\beta_{ik}$ is the local regression coefficient for the $k$th independent variable at location $i$; and $\epsilon_{i}$ is the random error at location $i$.

As data are geographically weighted, nearer observations have more influence in estimating the local set of regression coefficients than observations farther away. The model measures the inherent relationships around each regression point $i$, where each set of regression coefficients is estimated by a weighted least squares approach. The matrix expression for this estimation is:
\begin{equation*}
\hat{\beta}_i=\left(X^\top W(u_i,v_i)X\right)^{-1}X^\top W(u_i,v_i)y 
\end{equation*}
where $X$ is the matrix of the independent variables with a column of 1s for the intercept; $y$ is the dependent variable vector; $\hat{\beta}_i=(\beta_{i0},\ldots,\beta_{im})^\top $ is the vector of $m+1$ local regression coefficients; and $W_i$ is the diagonal matrix denoting the geographical weighting of each observed data for regression point $i$ at location $(u_i,v_i)$.  This weighting is determined by some kernel function as described in Section~\ref{sec.distbw}.

An optimum kernel bandwidth for GW regression can be found by minimising some model fit diagnostic, such as a leave-one-out cross-validation (CV) score \citep{Bowm:84a}, which only accounts for model prediction accuracy; or the Akaike Information Criterion (AIC) \citep{Akai:73a}, which accounts for model parsimony (i.e., a trade-off between prediction accuracy and complexity). In practice, a corrected version of the AIC is used, which unlike basic AIC is a function of sample size \citep{HurSim:98a}.  For GW regression, this entails that fits using small bandwidths receive a higher penalty (i.e., are more complex) than those using large bandwidths.  Thus for a GW regression with a bandwidth $b$, its AICc can be found from:

\begin{equation*}
\mathrm{AIC}_c(b)=2n\ln(\hat{\sigma})+n\ln(2\pi)+n\left\{\frac{n+\mathrm{tr}(S)}{n-2-\mathrm{tr}(S)} \right\}
\end{equation*}

where $n$ is the (local) sample size (according to $b$); $\hat{\sigma}$ is the estimated standard deviation of the error term; and $\mathrm{tr}(S)$ denotes the trace of the hat matrix $S$.  The hat matrix is the projection matrix from the observed $y$ to the fitted values, $\hat{y}$.

\subsection[Robust GW regression]{Robust GW regression}
To identify and reduce the effect of outliers in GW regression, various robust extensions have been proposed, two of which are described in \cite{FoBrCh:02a}. The first robust model re-fits a GW regression with a filtered data set that has been found by removing observations that correspond to large externally studentised residuals of an initial GW regression fit. An externally studentised residual for each regression location $i$ is defined as:
\begin{equation*}
r_i=\frac{e_i}{\hat{\sigma}_{-i}\sqrt{q_{ii}}}
\end{equation*}

where $e_i$ is the residual at location $i$;  $\hat{\sigma}_{-i}$ is a leave-one-out estimate of $\hat{\sigma}$; and $q_{ii}$ is the $i$th element of $(I-S)(I-S)^\top $. Observations are deemed outlying and filtered from the data if they have $|r_i|>3$.  The second robust model, iteratively down-weights observations that correspond to large residuals. This (non-geographical) weighting function $w_r$ on the residual $e_i$ is typically taken as:

\begin{equation*}
w_r(e_i) = \begin{cases} 1, & \mbox{if } |e_i|\leq 2\hat{\sigma} \\ 
\left[ 1-(|e_i|-2)^2\right]^2, & \mbox{if } 2\hat{\sigma}<|e_i|< 3\hat{\sigma}\\
0 & \mbox{otherwise}  \end{cases}
\end{equation*}

Observe that both approaches have an element of subjectivity, where the filtered data approach depends on the chosen residual cut-off (in this case, 3) and the iterative (automatic) approach depends on the chosen down-weighting function, with its associated cut-offs.

\subsection[Example]{Example}

We now demonstrate the fitting of the basic and robust GW regressions described, to the Dublin voter turnout data.  Our regressions attempt to accurately predict the proportion of the electorate who turned out on voting night to cast their vote in the 2004 General Election in Ireland.  The dependent variable is \code{GenEl2004} and the eight independent variables are \code{DiffAdd}, \code{LARent}, \code{SC1}, \code{Unempl}, \code{LowEduc}, \code{Age18_24}, \code{Age25_44} and \code{Age45_64}.

A global correlation analysis suggests that voter turnout is negatively associated with the independent variables, except for social class (\code{SC1}) and older adults (\code{Age45_64}).  Public renters (\code{LARent}) and unemployed (\code{Unempl}) have the highest correlations (both negative), in this respect.  The GW correlation analysis from Section 4 indicates that some of these relationships are non-stationary.  The global regression fit to this data yields an R-squared value of 0.63 and details of this fit can be summarised as follows:

\begin{CodeChunk}
\begin{CodeInput}
R> lm.global <- lm(GenEl2004 ~ DiffAdd + LARent + SC1 + Unempl + LowEduc + 
+ Age18_24 + Age25_44 + Age45_64, data = Dub.voter)
R> summary(lm.global)
\end{CodeInput}
\begin{CodeOutput}
   Coefficients:
               Estimate Std. Error t value Pr(>|t|)    
   (Intercept) 77.70467    3.93928  19.726  < 2e-16 ***
   DiffAdd     -0.08583    0.08594  -0.999   0.3187    
   LARent      -0.09402    0.01765  -5.326 1.92e-07 ***
   SC1          0.08637    0.07085   1.219   0.2238    
   Unempl      -0.72162    0.09387  -7.687 1.96e-13 ***
   LowEduc     -0.13073    0.43022  -0.304   0.7614    
   Age18_24    -0.13992    0.05480  -2.554   0.0111 *  
   Age25_44    -0.35365    0.07450  -4.747 3.15e-06 ***
   Age45_64    -0.09202    0.09023  -1.020   0.3086    
\end{CodeOutput}
\end{CodeChunk}

Next, we conduct a model specification exercise in order to help find an independent variable subset for our basic GW regression. As an aide to this task, a pseudo stepwise procedure is used that proceeds in a forward direction. The procedure can be described in the following four steps, where the results are visualised using associated plots of each model's $\mathrm{AIC}_c$ values:
\begin{itemize}
\item[Step 1.]	Start by calibrating all possible bivariate GW regressions by sequentially regressing a single independent variable against the dependent variable;
\item[Step 2.]	Find the best performing model which produces the minimum $\mathrm{AIC}_c$, and permanently include the corresponding independent variable in subsequent models;
\item[Step 3.]	Sequentially introduce a variable from the remaining group of independent variables to construct new models with the permanently included independent variables, and determine the next permanently included variable from the best fitting model that has the minimum $\mathrm{AIC}_c$; 
\item[Step 4.]	Repeat step 3 until all independent variables are permanently included in the model.
\end{itemize}

\begin{figure}
\centering
  \includegraphics[width=.8\linewidth]{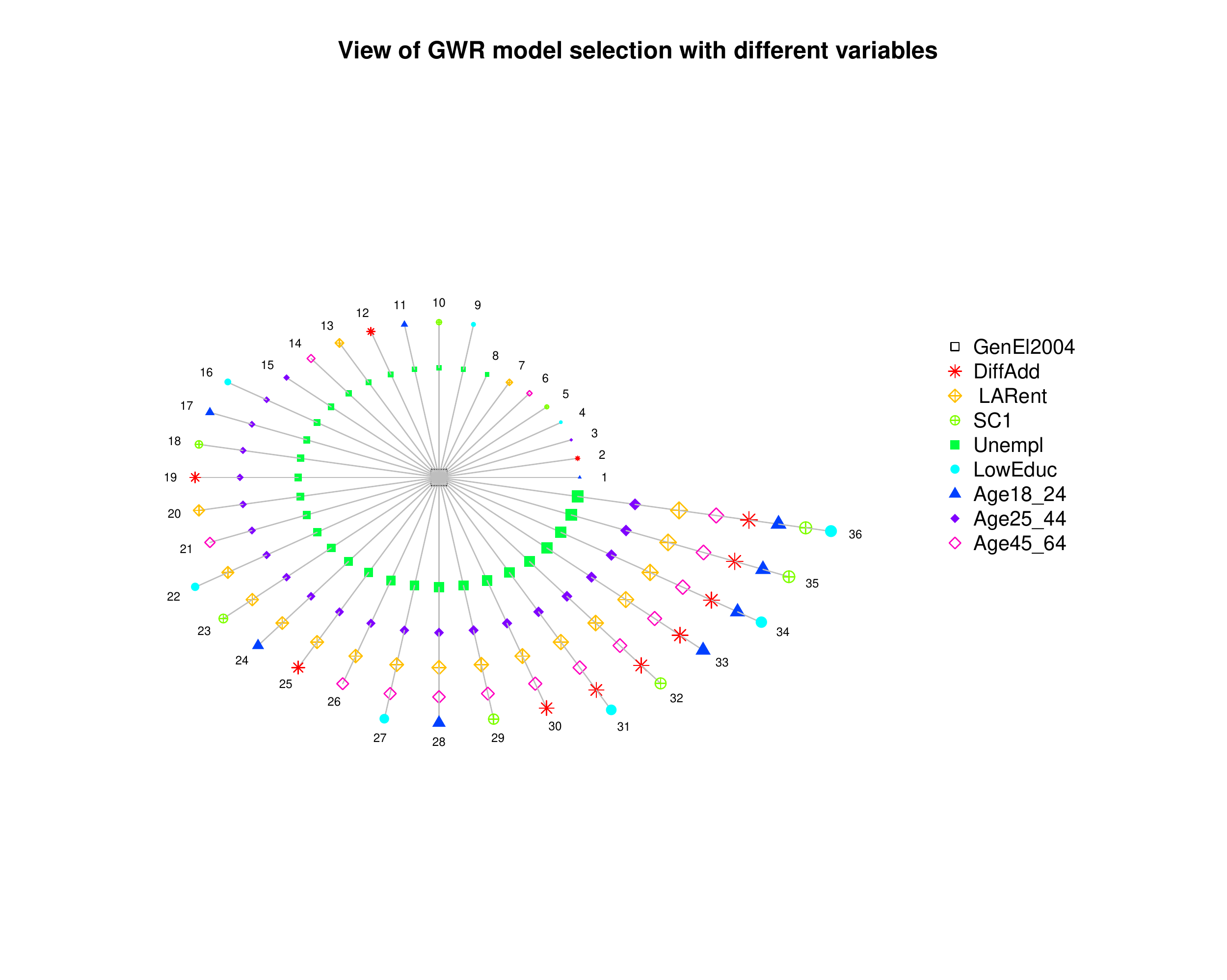}
  \caption{Model view of the stepwise specification procedure.}
  \label{fig:gwr_7}
\end{figure}

The function to perform this procedure is \code{model.selection.gwr}, whose AICc outputs are sorted using \code{model.sort.gwr} and then inputted to \code{model.view.gwr} to provide a useful visualisation of the $\mathrm{AIC}_c$ data (see Figure~\ref{fig:gwr_7}).  This approach can be conducted in a rudimentary form, where the bandwidth is user-specified beforehand and remains the same for each GW regression fit.  Alternatively, a more refined model specification exercise enables the re-calculation of an optimal bandwidth for each GW regression fit.  As a demonstration, a rudimentary specification is conducted, by running the following sequence of commands.  Observe that a bi-square kernel is specified with a user-specified adaptive bandwidth of $N=80$.

\begin{CodeChunk}
\begin{CodeInput}
R> DeVar <- "GenEl2004"
R> InDeVars <- c("DiffAdd"," LARent", "SC1", "Unempl", "LowEduc",
+ "Age18_24", "Age25_44", "Age45_64")

R> model.sel <- model.selection.gwr(DeVar ,InDeVars, data = Dub.voter, 
+ kernel = "bisquare", adaptive = TRUE, bw = 80)
R> sorted.models <- model.sort.gwr(model.sel, numVars = length(InDeVars),
+ ruler.vector = model.sel[[2]][,2])
R> model.list <- sorted.models[[1]]

R> X11(width = 10, height = 8)
R> model.view.gwr(DeVar, InDeVars, model.list = model.list)

R> X11(width = 10, height = 8)
R> plot(sorted.models[[2]][,2], col = "black", pch = 20, lty = 5,
+ main = "Alternative view of GWR model selection procedure", ylab = "AICc",
+ xlab = "Model number", type = "b")
\end{CodeInput}
\end{CodeChunk}

\begin{figure}
\centering
  \includegraphics[width=.8\linewidth]{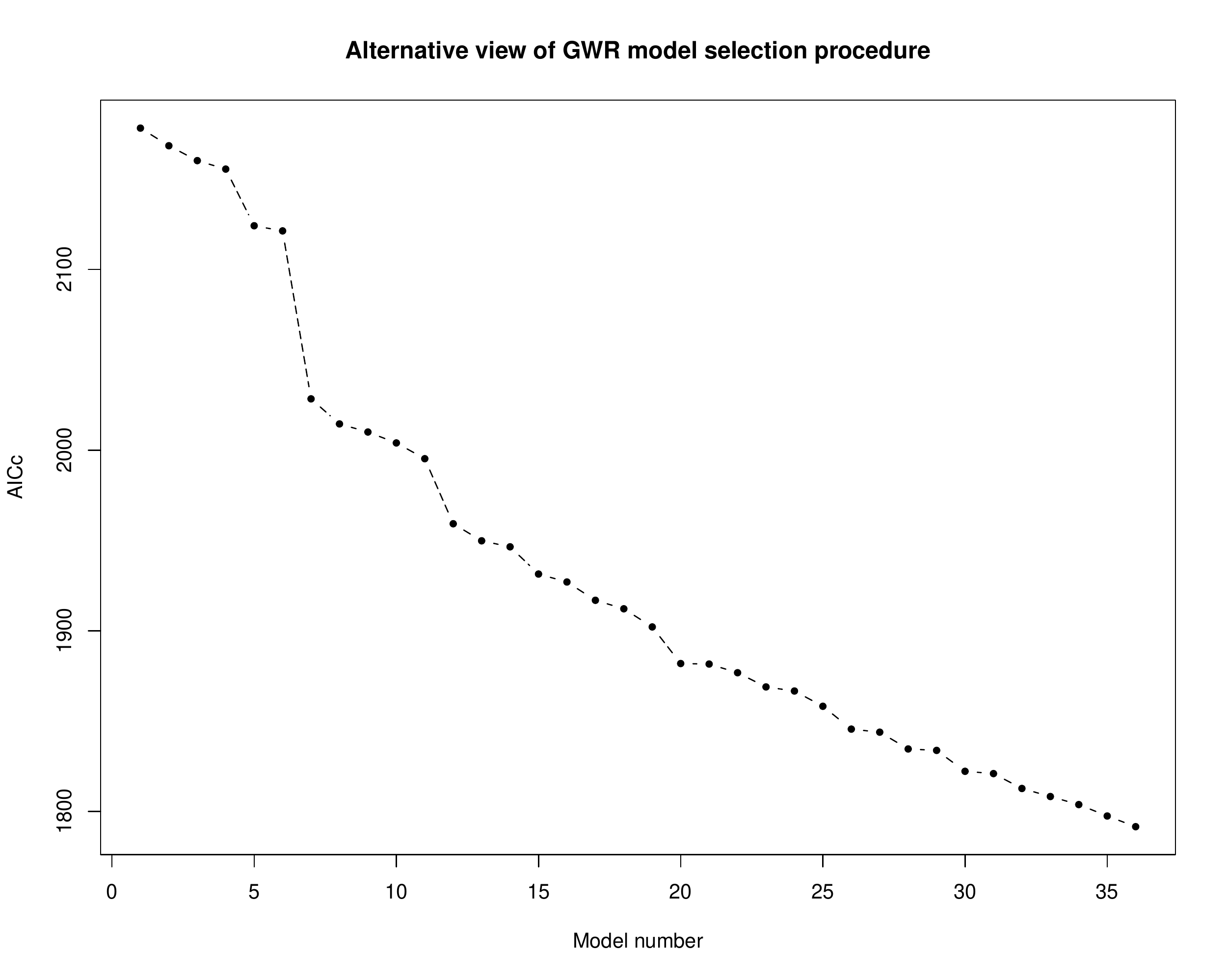}
  \caption{$\mathrm{AIC}_c$ values for the same 36 GW regressions of Figure~\ref{fig:gwr_7}.}
  \label{fig:gwr_8}
\end{figure}

Figure~\ref{fig:gwr_7} presents a circle view of the 36 GW regressions (numbered 1 to 36) that result from this step-wise procedure. Here the dependent variable is located in the centre of the chart and the independent variables are represented as nodes differentiated by shapes and colours. The first independent variable that is permanently included is \code{Unempl}, the second is \code{Age25_44}, and the last is \code{LowEduc}.  Figure~\ref{fig:gwr_8} displays the corresponding $\mathrm{AIC}_c$ values from the same fits of Figure~\ref{fig:gwr_7}.  The two graphs work together, explaining model performance when more and more variables are introduced.  Clearly, $\mathrm{AIC}_c$ values continue to fall until all independent variables are included.  Results suggest that continuing with all eight independent variables is worthwhile (at least for our user-specified bandwidth).

We can now proceed to the correct calibration of our chosen GW regression specification.  Here, we find its true (i.e., optimal) bandwidth using the function \code{bw.gwr} and then use this bandwidth to parametrise the same GW regression with the function \code{gwr.basic}.  The optimal bandwidth is found at $N=109$.  Commands for these operations are as follows, where the print function provides a useful report of the global and GW regression fits, with summaries of their regression coefficients, diagnostic information and F-test results \citep[following][]{LeMeZh:00a}.  The report is designed to match the output of the \proglang{GW regression v3.0} executable software \cite{gwr3}.

\begin{CodeChunk}
\begin{CodeInput}
R> bw.gwr.1 <- bw.gwr(GenEl2004 ~ DiffAdd + LARent + SC1 + Unempl + LowEduc +
+ Age18_24 + Age25_44 + Age45_64, data = Dub.voter, approach = "AICc", 
+ kernel = "bisquare", adaptive = TRUE)
R> bw.gwr.1
\end{CodeInput}
\begin{CodeOutput}
[1] 109
\end{CodeOutput}
\begin{CodeInput}
R> gwr.res <- gwr.basic(GenEl2004 ~ DiffAdd + LARent + SC1 + Unempl + 
+ LowEduc + Age18_24 + Age25_44 + Age45_64, data = Dub.voter, 
+ bw = bw.gwr.1,  kernel = "bisquare", adaptive = TRUE, F123.test = TRUE)

R> print(gwr.res)
\end{CodeInput}
\end{CodeChunk}

\begin{figure}
\centering
\begin{subfigure}{.5\textwidth}
  \centering
  \includegraphics[width=1\linewidth]{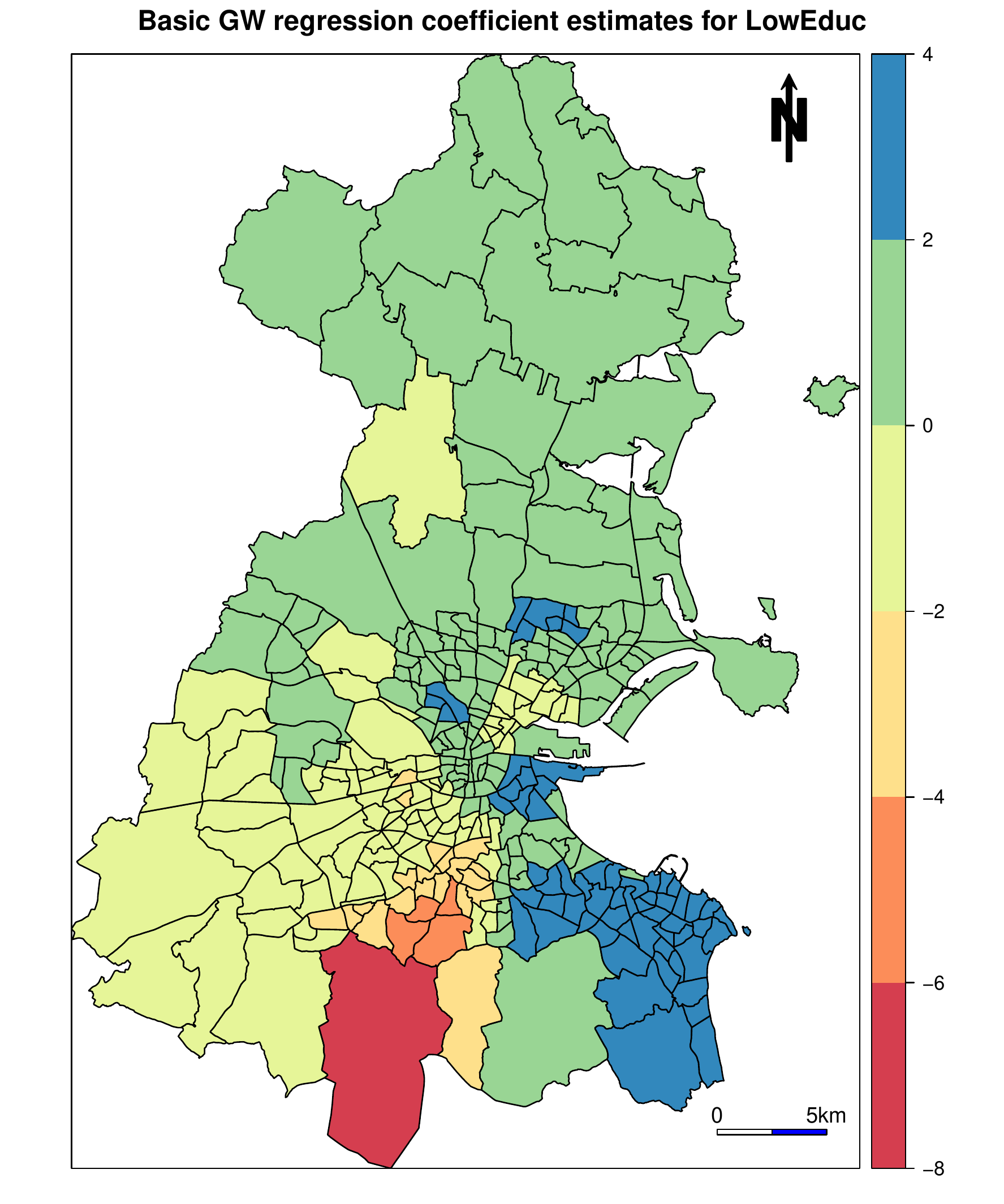}
  \caption{}
  \label{fig:gwr_9a}
\end{subfigure}%
\begin{subfigure}{.5\textwidth}
  \centering
  \includegraphics[width=1\linewidth]{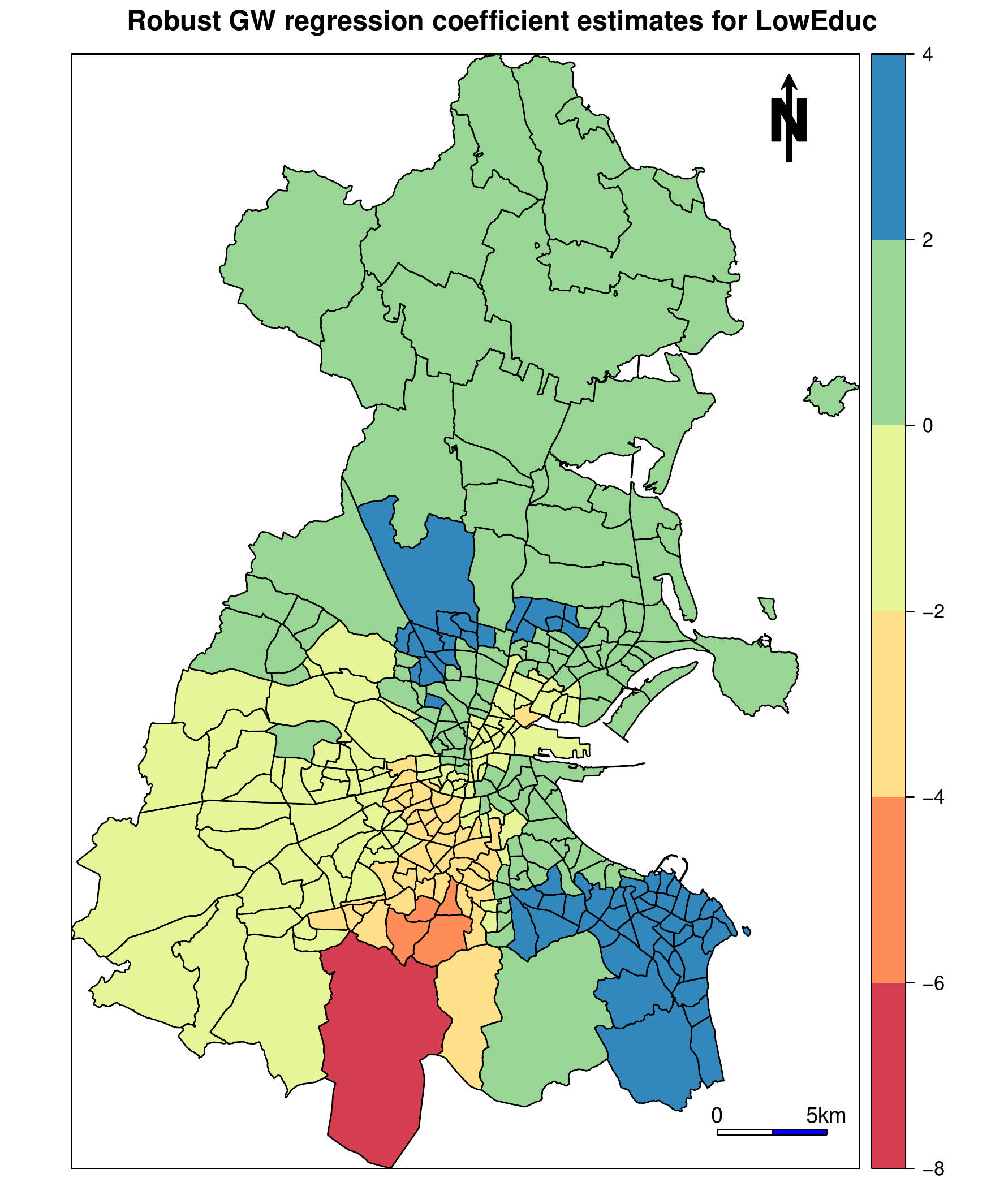}
  \caption{}
  \label{fig:gwr_9b}
\end{subfigure}
\caption{(a) Basic and (b) robust GW regression coefficient estimates for \code{LowEduc}.}
\label{fig:gwr9}
\end{figure}

To map the GW regression outputs, the following commands can be used to each field of spatial data frame object \code{gwr.res$SDF}.  As an example, we map the coefficient estimates for \code{LowEduc} in Figure~\ref{fig:gwr_9a}, where this variable's relationship to voter turnout has clear geographical variation, ranging from -7.67 to 3.41.  Its global regression coefficient estimate is -0.13.  Commands for a robust GW regression fit (the second, iterative approach) of the same model, using the same bandwidth, are also given.  Here a slightly different set of coefficient estimates for \code{LowEduc} result (Figure~\ref{fig:gwr_9b}), to that found with the basic fit.  Evidence for relationship non-stationarity is now slightly weaker, as the robustly estimated coefficients range from -7.74 to 2.57, but the broad spatial pattern in these estimates remain largely the same.

\begin{CodeChunk}
\begin{CodeInput}
R> names(gwr.res$SDF) 
R> mypalette.6 <- brewer.pal(6, "Spectral")

R> X11(width=10,height=12)
R> spplot(gwr.res$SDF, "LowEduc", key.space = "right", 
+ col.regions = mypalette.6, at = c(-8, -6, -4, -2, 0, 2, 4), 
+ main = "Basic GW regression coefficient estimates for LowEduc",
+ sp.layout=map.layout)

R> rgwr.res <- gwr.robust(GenEl2004 ~ DiffAdd + LARent + SC1 + Unempl + 
+ LowEduc + Age18_24 + Age25_44 + Age45_64, data = Dub.voter, bw = bw.gwr.1,
+ kernel = "bisquare", adaptive = TRUE, F123.test = TRUE)

R> print(rgwr.res)

R> X11(width = 10, height = 12)
R> spplot(rgwr.res$SDF, "LowEduc", key.space = "right",
+ col.regions = mypalette.6, at = c(-8, -6, -4, -2, 0, 2, 4), 
+ main = "Robust GW regression coefficient estimates for LowEduc",
+ sp.layout=map.layout)
\end{CodeInput}
\end{CodeChunk}

\section[GW regression and addressing local collinearity]{GW regression and addressing local collinearity}\label{sec.multicoll}
\subsection[Introduction]{Introduction}
A problem which has long been acknowledged in regression modelling is that of collinearity among the predictor (independent) variables.  The effects of collinearity include a loss of precision and a loss of power in the coefficient estimates. Collinearity is potentially more of an issue in GW regression because: (i) its effects can be more pronounced with the smaller spatial samples used in each local estimation and (ii) if the data are spatially heterogeneous in terms of its correlation structure, some localities may exhibit collinearity while others may not.  In both cases, collinearity may be a source of problems in GW regression even when no evidence is found for collinearity in the global model \citep{WheTie:05a, whe07, whe13}. A further complication is that in the case of a predictor which has little local spatial variation, the possibility of collinearity with the intercept term is raised \citep{whe07, whe10, whe13}. Simulation studies have indicated that in the presence of collinearity, GW regression may find patterns in the coefficients where no patterns are actually present \citep{WheTie:05a,PaFaWe11}.

To this extent, diagnostics to investigate the nature of collinearity in a GW regression analysis should always be conducted; this includes finding: (a) local correlations amongst pairs of predictors; (b) local variance inflation factors (VIFs) for each predictor; (c) local variance decomposition proportions (VDPs); and (d) local design (or cross-product) matrix condition numbers; all at the same spatial scale of each local regression of the GW regression model. Accordingly, the following rules of thumb can be taken to indicate likely local collinearity problems in the GW regression fit: (a) absolute local correlations greater than 0.8 for a given predictor variable pair; (b) VIFs greater than 10 for a given predictor; (c) VDPs greater than 0.5; and (d) condition numbers greater than 30.  Such diagnostics and associated rules of thumb are directly taken from the global regression case \citep{bekuWe80,obr07} and have been proposed in a GW regression context through the works of \cite{WheTie:05a, whe07}.  All four diagnostics can be found and mapped using the function \code{gwr.collin.diagno} in \pkg{GWmodel} and a similar function exists in the \pkg{gwrr} \proglang{R} package, see Section~\ref{sec.gwcomp}. Here it should be noted that local correlations and local VIFs cannot detect collinearity with the intercept; \cite{whe10} provides a useful example of this.  To this extent, the combined use of local VDPs and local condition numbers are superior diagnostics to local correlations and local VIFs for investigating collinearity \citep{whe07}.

There are several possible actions in the light of discovering high levels of collinearity.  These include: (1) doing nothing, (2) removing the offending predictors, (3) transforming the predictors to some orthogonal form or (4) using a different non-stationary regression model.  The removal of a predictor is recommended if it clearly creates a global and local collinearity issue.  However removing a predictor is not ideal, when only a local collinearity effect is present.  Transforming the predictors should be done locally, at the same spatial scale of the GW regression (e.g. with a GW PCA), rather than globally (with PCA).  However, such operations entail a loss of meaning in the regression outputs.  For local collinearity issues, the fourth option is the best option and here \cite{whe07,whe09} proposed significant modifications to the GW regression model, designed to cater for the adverse effects of collinearity.  In particular, penalised regression models were transferred to a local form with the GW ridge regression model \citep{whe07,whe09} and the GW lasso \citep{whe09}.  Both penalised GW models are biased estimators, providing no measures of coefficient uncertainty, but the GW lasso has the advantage in that it can also provide a local model selection function. In the presence of collinearity, both penalised GW models should provide more accurate local coefficient estimates, than that found with basic GW regression.  Thus an investigation of relationship non-stationary should be more assured in this respect.

In this section, our aim is to present the use of local condition numbers as a diagnostic for local collinearity and also to relate this diagnostic to the ridge parameter of a GW ridge regression model.  This relationship enables us to provide an alternative GW ridge regression to that demonstrated in \cite{whe07}.  We call our new model, GW regression with a locally-compensated ridge term.  This model differs to any existing GW ridge regression, in that: (A) it fits local ridge regressions with their own ridge parameters (i.e., the ridge parameter varies across space) and (B) it only fits such ridge regressions at locations where the local condition number is above a user-specified threshold.  Thus a biased local estimation is not necessarily used everywhere; only at locations where collinearity is likely to be an issue.  At all other locations, the usual un-biased estimator is used. \pkg{GWmodel} functions that we demonstrate are \code{bw.gwr.lcr}, to optimally estimate the bandwidth, and \code{gwr.lcr}, to fit the locally-compensated GW regression.    

\subsection[Ridge regression]{Ridge regression}
A method to reduce the adverse effects of collinearity in the predictors of a linear model is ridge regression \citep{hoe62,hoeken70}.  Other methods include principal components regression and partial least squares regression \citep{FrFr93}.  In ridge regression the estimator is altered to include a small change to the values of the diagonal of the cross-product matrix. This is known as the ridge, indicated by $\lambda$ in the following equation:
\begin{equation*}
\hat{\beta}=\left(X^\top X+\lambda I \right)^{-1}X^\top Y
\end{equation*}

The effect of the ridge is to increase the difference between the diagonal elements of the matrix and the off-diagonal elements. As the off-diagonal elements represent the co-variation in the predictors, the effect of the collinearity among the predictors in the estimation is lessened.  The price of this is that $\hat{\beta}$ becomes biased, and the standard errors (and associated $t$-values) of the estimates are no longer available. Of interest is the value to be given to the ridge parameter; \cite{lee87} presents an algorithm to find a value which yields the best predictions. 

\subsection[GW regression with local compensation]{GW regression with local compensation}\label{sec.lcrgwr}

There exists a link between the definition of the condition number for the cross-product matrix ($X^\top X$) and the ridge parameter based on the observation that if the eigenvalues of $X^\top X$ are $\epsilon_1,\epsilon_2,\ldots,\epsilon_p $ then the eigenvalues of $X^\top X+\lambda I$ are $\epsilon_1+\lambda,\epsilon_2+\lambda,\ldots,\epsilon_p+\lambda$. The condition number $\kappa$ of a square matrix is defined as $\epsilon_1/\epsilon_p$, so the condition number for the ridge-adjusted matrix will be $\epsilon_1+\lambda/\epsilon_p+\lambda$.  By re-arranging the terms, the ridge adjustment that will be required to yield a particular condition number $\kappa$ is $\lambda=\left\{(\epsilon_1-\epsilon_p)/(\kappa-1)\right\}-\epsilon_p$.  Thus given the eigenvalues of the un-adjusted matrix, and the desired condition number, we can determine the value of the ridge which is required to yield that condition number.

For GW regression, this can be applied to the GW cross-product matrix, which permits a \emph{local compensation} of each local regression model, so that the local condition number never exceeds a specified value of $\kappa$.  The condition numbers for the un-adjusted matrices may also be mapped to give an indication of where the analyst should take care in interpreting the results, or the local ridge parameters may also be mapped. For cases where collinearity is as much an issue in the global regression as in the GW regression; the local estimations will indicate precisely where the collinearity is a problem. The estimator for this locally compensated ridge (LCR) GW regression model is:

\begin{equation*}
\hat{\beta}(u_i,v_i)=\left(X^\top W(u_i,v_i)X+\lambda I(u_i,v_i) \right)^{-1}X^\top W(u_i,v_i)Y
\end{equation*}

where $\lambda I(u_i,v_i)$ is the locally-compensated value of $\lambda$ at location $(u_i,v_i)$.  Observe that the same approach to estimating the bandwidth in the basic GW regression (Section~\ref{sec.gwr}) can be applied to the LCR GW regression model. For a cross-validation approach, the bandwidth is optimised to yield the best predictions. Collinearity tends to have a greater affect on the coefficient estimates rather than the predictions from the model, so in general, little is lost when using the locally-compensated form of the model.  Details on this and an alternative locally-compensated GW regression can be found in \cite{Coll2012}, where both models are performance tested within a simulation experiment.

\subsection[Example]{Example}

We examine the use of our local compensation approach with the same GW regression that is specified in Section~\ref{sec.gwr}, where voter turnout is a function of the eight predictor variables of the Dublin election data.  For the corresponding global regression, the \code{vif} function in the \pkg{car} library computes VIFs using the method outlined in \cite{FoMo92}.  These global VIFs are given below and (noting their drawbacks given above) suggest that weak collinearity exists within this data.

\begin{CodeChunk}
\begin{CodeInput}
R> library("car")

R> lm.global <- lm(GenEl2004 ~ DiffAdd + LARent + SC1 + Unempl +
+ LowEduc + Age18_24 + Age25_44 + Age45_64, data=Dub.voter)
R> summary(lm.global)
R> vif(lm.global)
\end{CodeInput}
\begin{CodeOutput}
 DiffAdd   LARent      SC1   Unempl  LowEduc Age18_24 Age25_44 Age45_64 
3.170044 2.167172 2.161348 2.804576 1.113033 1.259760 2.879022 2.434470 
\end{CodeOutput}
\end{CodeChunk}

In addition, the PCA from Section~\ref{sec.gwpca} suggests collinearity between \code{DiffAdd}, \code{LARent}, \code{Unempl}, \code{Age25_44}, and \code{Age45_64}.  As the first component accounts for some 36\% of the variance in the data set, and of those components with eigenvalues greater than 1, the proportion of variance accounted for is 73.6\%, we might consider removing variables with higher loadings.  However for the purposes of illustration, we decide to keep the model as it is.  Further global findings are of note, in that the correlation of turnout with \code{Age45_64} is positive, but the sign of the global regression coefficient is negative. Furthermore, only four of the global regression predictors are significant.  Unexpected sign changes and relatively few significant variables are both indications of collinearity.

We can measure the condition number of the design matrix using the method outlined in \cite{bekuWe80}.  The method, termed BKW, requires that the columns of the matrix are scaled to have length 1; the condition number is the ratio of the largest to the smallest singular value of this matrix. The following code implements the BKW computations, where $X$ is the design matrix consisting of the predictor variables and a column of 1s.

\begin{CodeChunk}
\begin{CodeInput}
R> X <- as.matrix(cbind(1,Dub.voter@data[,4:11]))
R> BKWcn <- function(X) {
+    p <- dim(X)[2]
+    Xscale <- sweep(X, 2, sqrt(colSums(X^2)), "/")
+    Xsvd <- svd(Xscale)$d
+    Xsvd[1] / Xsvd[p]
+    }
R> BKWcn(X)
\end{CodeInput}
\begin{CodeOutput}
[1] 41.06816
\end{CodeOutput}
\end{CodeChunk}

The BKW condition number is found to be 41.07 which is high, indicating that collinearity is at least, a global problem for this data.  We can experiment by removing columns from the design matrix and test which variables appear to be the source of the collinearity. For example, entering:

\begin{CodeChunk}
\begin{CodeInput}
R> BKWcn(X[,c(-2,-8)])
\end{CodeInput}
\begin{CodeOutput}
[1] 18.69237
\end{CodeOutput}
\end{CodeChunk}

allows us to examine the effects of removing both \code{DiffAdd} and \code{Age25_44} as sources of collinearity.   The reduction of the BKW condition number to 18.69 suggests that removing these two variables is a useful start.  However, for demonstration purposes, we will persevere with the collinear (full specification) model, and now re-examine its GW regression fit, the one already fitted in Section~\ref{sec.gwr}. The main function to perform this collinearity assessment is \code{gwr.lcr}, where we aim to compare the coefficient estimates for the un-adjusted basic GW regression with those from a LCR GW regression.

In the first instance, we can use this function to find the global condition number (as that found with the global regression).  This can be done simply by specifying a box-car kernel with a bandwidth equal to the sample size.  This is equivalent to fitting $n$ global models.  Inspection of the results from the spatial data frame show that the condition numbers are all equal to 41.07, as hoped for.  The same condition number is outputted by the \proglang{ArcGIS} Geographically weighted Regression tool in the Spatial Statistics Toolbox \citep{arcgis}.  Commands to conduct this check on the behaviour of the \code{lcr.gwr} function are as follows:

\begin{CodeChunk}
\begin{CodeInput}
R> nobs <- dim(Dub.voter)[1]
R> lcrm1 <- gwr.lcr(GenEl2004 ~ DiffAdd + LARent + SC1 + Unempl + LowEduc + 
+ Age18_24 + Age25_44 + Age45_64, data = Dub.voter, bw = nobs, 
+ kernel = "boxcar", adaptive=TRUE)
R> summary(lcrm1$SDF$Local_CN)
\end{CodeInput}
\begin{CodeOutput}
   Min. 1st Qu.  Median    Mean 3rd Qu.    Max. 
  41.07   41.07   41.07   41.07   41.07   41.07 
\end{CodeOutput}
\end{CodeChunk}

To obtain local condition numbers for a basic GW regression without a local compensation, we use the \code{bw.gwr.lcr} function to optimally estimate the bandwidth, and then \code{gwr.lcr} to estimate the local regression coefficients and the local condition numbers\footnote{Local condition numbers can also be found using the  function \code{gwr.collin.diagno}.}.  To match that of Section~\ref{sec.gwr}, we specify an adaptive bi-square kernel.  Observe that the bandwidth for this model can be exactly the same as that obtained using \code{bw.gwr}, the basic bandwidth function. With no local compensation (i.e., local ridges of zero), the cross-products matrices will be identical, but only provided the same optimisation approach is specified.  Here we specify a cross-validation (CV) approach, as the AICc approach is currently not an option in the \code{bw.gwr.lcr} function.  Coincidently, for our basic GW regression of Section~\ref{sec.gwr}, a bandwidth of $N=109$ results for both CV and AICc approaches.  Commands to output the local condition numbers from our basic GW regression, and associated model comparisons are as follows:

\begin{CodeChunk}
\begin{CodeInput}
R> lcrm2.bw <- bw.gwr.lcr(GenEl2004 ~ DiffAdd + LARent + SC1 + Unempl + 
+ LowEduc + Age18_24 + Age25_44 + Age45_64, data = Dub.voter, 
+ kernel = "bisquare", adaptive=TRUE)
R> lcrm2.bw
\end{CodeInput}
\begin{CodeOutput}
[1] 109
\end{CodeOutput}
\begin{CodeInput}
R> lcrm2 <- gwr.lcr(GenEl2004 ~ DiffAdd + LARent + SC1 + Unempl + LowEduc + 
+ Age18_24 + Age25_44 + Age45_64, data = Dub.voter, bw = lcrm2.bw, 
+ kernel = "bisquare", adaptive = TRUE)
R> summary(lcrm2$SDF$Local_CN)
\end{CodeInput}
\begin{CodeOutput}
   Min. 1st Qu.  Median    Mean 3rd Qu.    Max. 
  32.88   52.75   59.47   59.28   64.85  107.50 
\end{CodeOutput}
\begin{CodeInput}
R> gwr.cv.bw <- bw.gwr(GenEl2004 ~ DiffAdd + LARent + SC1 + Unempl + 
+ LowEduc + Age18_24 + Age25_44 + Age45_64, data = Dub.voter, 
+ approach = "CV", kernel = "bisquare", adaptive = TRUE)
R> gwr.cv.bw
\end{CodeInput}
\begin{CodeOutput}
[1] 109
\end{CodeOutput}
\begin{CodeInput}
R> mypalette.7 <- brewer.pal(8, "Reds")

R> X11(width = 10, height = 12)
R> spplot(lcrm2$SDF, "Local_CN", key.space = "right",
+ col.regions = mypalette.7, cuts=7,
+ main="Local condition numbers from basic GW regression",
+ sp.layout=map.layout)
\end{CodeInput}
\end{CodeChunk}

\begin{figure}
\centering
\begin{subfigure}{.5\textwidth}
  \centering
  \includegraphics[width=1\linewidth]{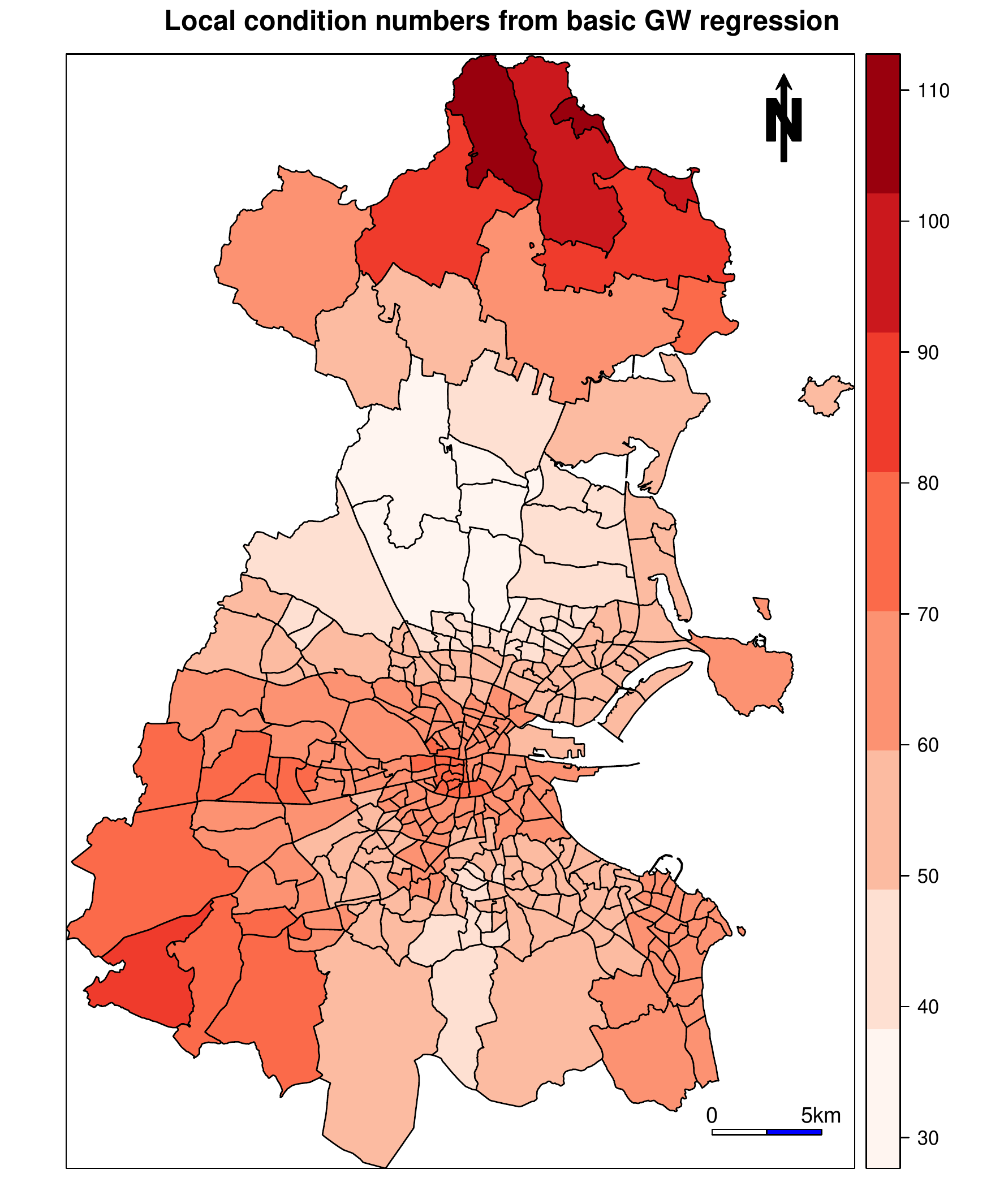}
  \caption{}
  \label{fig:gwcoll_10a}
\end{subfigure}%
\begin{subfigure}{.5\textwidth}
  \centering
  \includegraphics[width=1\linewidth]{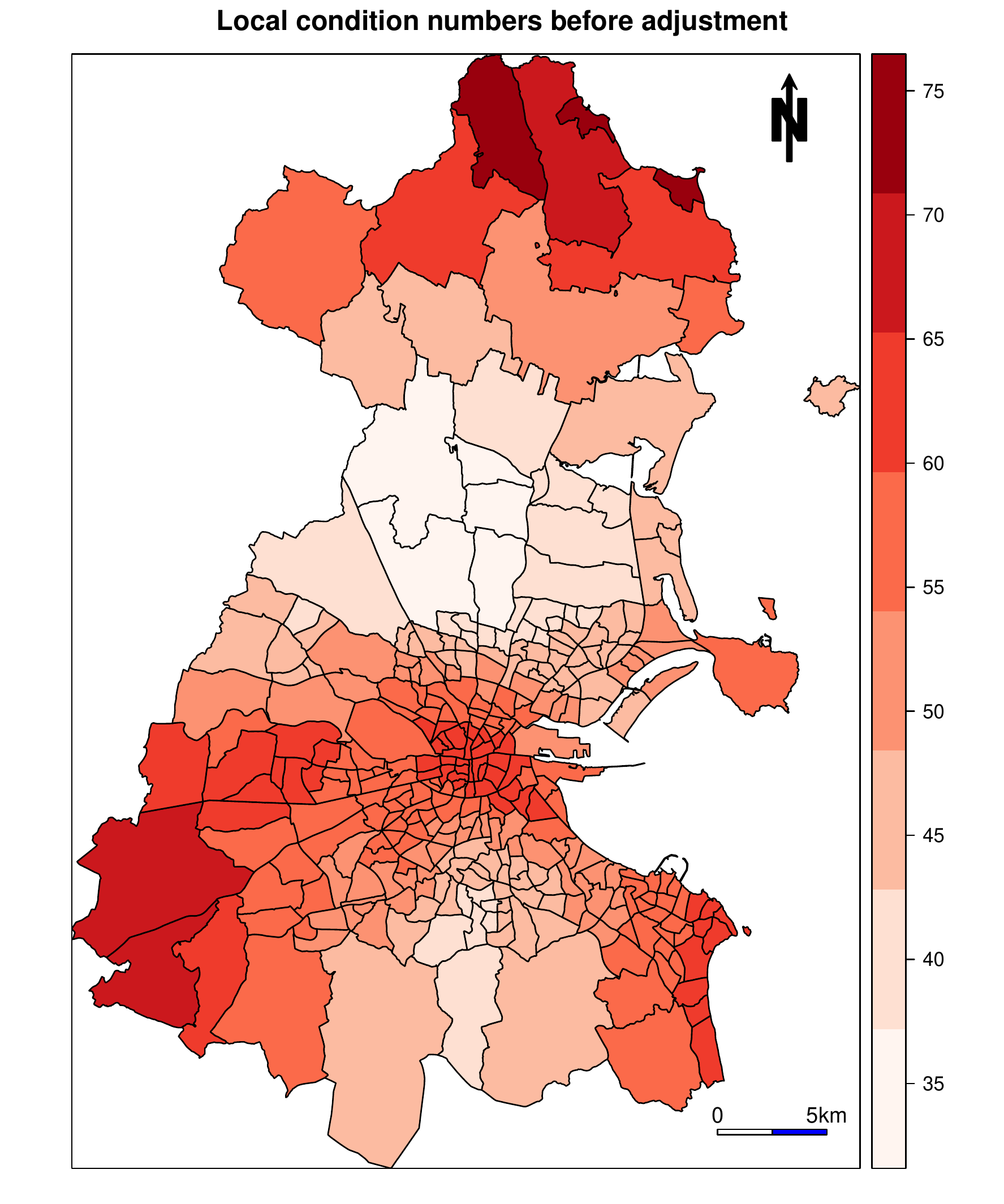}
  \caption{}
  \label{fig:gwcoll_10b}
\end{subfigure}
\caption{Local condition numbers from: (a) basic GW regression and (b) before adjustment.}
\label{fig:gwcoll10}
\end{figure}

Thus the local condition numbers can range from 32.88 to 107.50, all worryingly large everywhere.  Whilst the local estimations are potentially more susceptible to collinearity than the global model, we might consider removing some of the variables which cause problems globally.  The maps will show where the problem is worst, and where action should be concentrated.  The local condition numbers for this estimation are shown in Figure~\ref{fig:gwcoll_10a}.

We can now calibrate a LCR GW regression, where the application of local ridge adjustment to each local $X^\top W(u_i,v_i)X$ matrix, only occurs at locations where the local condition number exceeds some user-specified threshold. At these locations, the local compensation forces the condition numbers not to exceed the same specified threshold. In this case, we follow convention and specify a threshold of 30. As a consquence, a local ridge term will be found at all locations, for this particular example. The \code{lambda.adjust = TRUE} and \code{cn.thresh = 30} arguements in the \code{gwr.lcr} function are used to invoke the local compensation process, as can be seen in following commands:

\begin{CodeChunk}
\begin{CodeInput}
R> lcrm3.bw <- bw.gwr.lcr(GenEl2004 ~ DiffAdd + LARent + SC1 + Unempl + 
+ LowEduc + Age18_24 + Age25_44 + Age45_64, data = Dub.voter, 
+ kernel = "bisquare", adaptive = TRUE, lambda.adjust = TRUE, cn.thresh = 30)
R> lcrm3.bw
\end{CodeInput}
\begin{CodeOutput}
[1] 157
\end{CodeOutput}
\begin{CodeInput}
R> lcrm3 <- gwr.lcr(GenEl2004 ~ DiffAdd + LARent + SC1+ Unempl + LowEduc + 
+ Age18_24 + Age25_44 + Age45_64, data=Dub.voter, bw = lcrm3.bw, 
+ kernel = "bisquare", adaptive = TRUE, lambda.adjust = TRUE, cn.thresh = 30)
R> summary(lcrm3$SDF$Local_CN)
\end{CodeInput}
\begin{CodeOutput}
   Min. 1st Qu.  Median    Mean 3rd Qu.    Max. 
  34.34   47.08   53.84   52.81   58.66   73.72 
\end{CodeOutput}
\begin{CodeInput}
R> summary(lcrm3$SDF$Local_Lambda)
\end{CodeInput}
\begin{CodeOutput}
   Min. 1st Qu.  Median    Mean 3rd Qu.    Max. 
0.01108 0.03284 0.04038 0.03859 0.04506 0.05374 
\end{CodeOutput}
\begin{CodeInput}
R> X11(width = 10, height = 12)
R> spplot(lcrm3$SDF, "Local_CN", key.space = "right",
+ col.regions = mypalette.7, cuts = 7,
+ main = "Local condition numbers before adjustment",
+ sp.layout=map.layout)

R> X11(width = 10, height = 12)
R> spplot(lcrm3$SDF, "Local_Lambda", key.space = "right",
+ col.regions = mypalette.7,cuts = 7,
+ main = "Local ridge terms for LCR GW regression",
+ sp.layout=map.layout)
\end{CodeInput}
\end{CodeChunk}

Observe that the bandwidth for the locally compensated GW regression is larger at  $N=157$, than for the un-adjusted (basic) GW regression (at $N=109$).  We could have specified the bandwidth from the un-adjusted model, but this would not provide the best fit.  The larger bandwidth provides greater smoothing.  Observe also that the local condition numbers (the \code{Local_CN} outputs) from this model are the local condition numbers \emph{before} the adjustment (or compensation)\footnote{\emph{After} the adjustment, the local condition numbers will all equal 30 in this case.}.  They will tend to be smaller than those for the basic model because we are using a larger bandwidth. They will tend to that of the global model (i.e., 41.07).

\begin{figure}
\centering
\begin{subfigure}{.5\textwidth}
  \centering
  \includegraphics[width=1\linewidth]{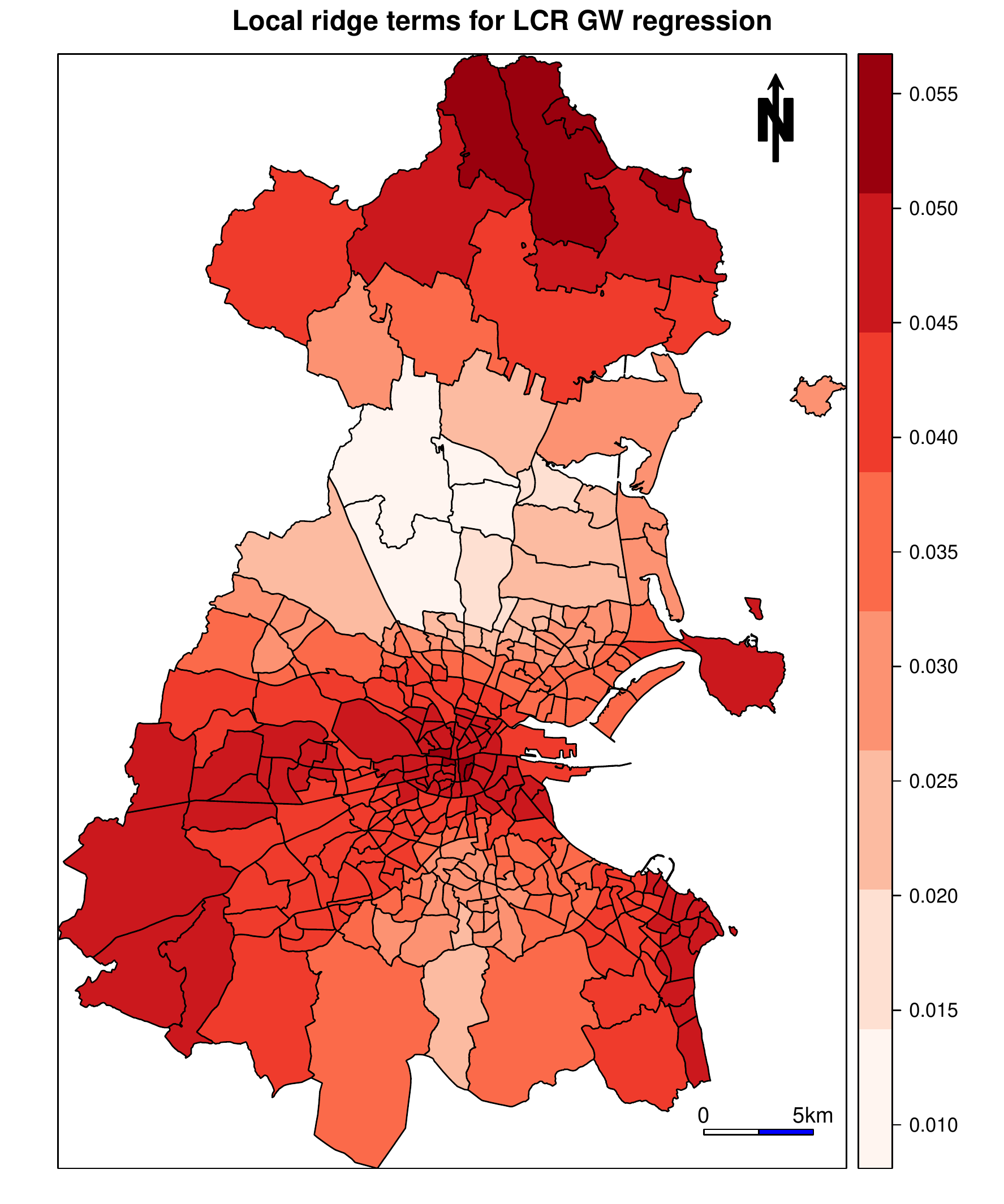}
  \caption{}
  \label{fig:gwcoll_11a}
\end{subfigure}%
\begin{subfigure}{.5\textwidth}
  \centering
  \includegraphics[width=1\linewidth]{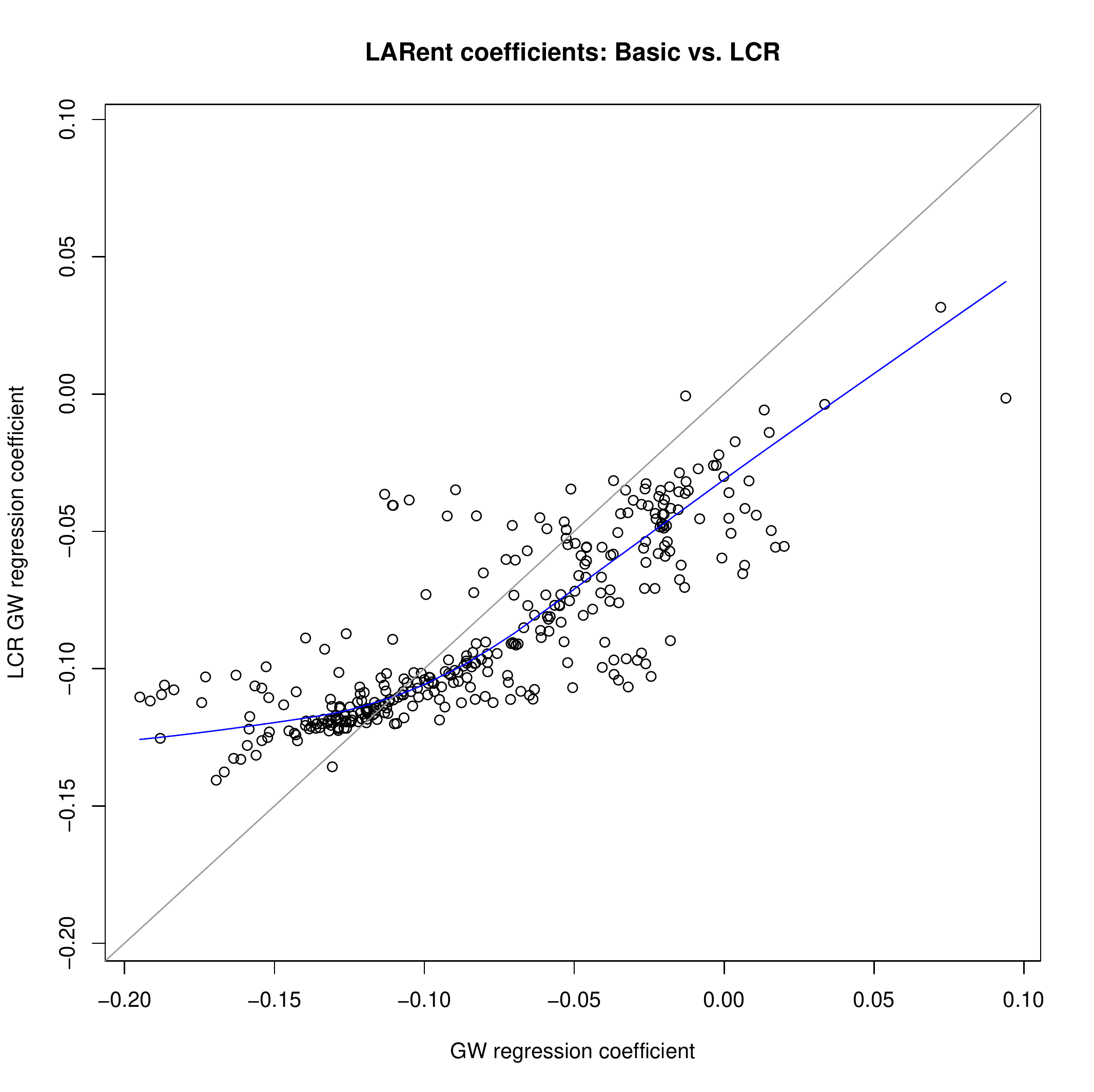}
  \caption{}
  \label{fig:gwcoll_11b}
\end{subfigure}
\caption{(a) Local ridge terms and (b) comparison of coefficient estimates for \code{LARent}.}
\label{fig:gwcoll11}
\end{figure}

The \code{Local_Lambda} outputs are the local ridge estimates used to adjust the local cross-products matrices.  Both the local condition numbers and the local ridges can be mapped to show where the GW regression has applied different levels of adjustment in relation to the different levels of collinearity among the predictors.  The local condition numbers are mapped in Figure~\ref{fig:gwcoll_10b}, and the local ridges in Figure~\ref{fig:gwcoll_11a} for our LCR GW regression. The greatest adjustments were required in central Dublin, and in the north, south-west and south-east extremities of the study area.

Figure~\ref{fig:gwcoll_11b} plots the adjusted coefficient estimates for \code{LARent} from the locally compensated model, against those from the corresponding basic model.  The general pattern would appear to be that the larger coefficients for the basic model are reduced in magnitude, and that the smaller coefficients are raised. The relationship is non-linear and a loess fit is shown in the plot. Commands for this comparison are as follows:

\begin{CodeChunk}
\begin{CodeInput}
R> gwr.cv <- gwr.basic(GenEl2004 ~ DiffAdd + LARent + SC1 + Unempl +
+ LowEduc + Age18_24 + Age25_44 + Age45_64, data = Dub.voter, bw = gwr.cv.bw, 
+ kernel = "bisquare", adaptive = TRUE)

R> small <- min(min(gwr.cv$SDF$LARent), min(lcrm3$SDF$LARent))
R> large <- max(max(gwr.cv$SDF$LARent), max(lcrm3$SDF$LARent))

R> X11(w=10,h=10)
R> plot(gwr.cv$SDF$LARent,lcrm3$SDF$LARent,
+ main = " LARent coefficients: basic vs. locally compensated",
+ xlab = "GW regression coefficient", ylab = "LCR GW regression coefficient",
+ xlim = c(small, large), ylim = c(small, large))
R> lines(lowess(gwr.cv$SDF$LARent, lcrm3$SDF$LARent), col = "blue")
R> abline(0, 1, col = "gray60")
\end{CodeInput}
\end{CodeChunk}

\subsubsection[Model building with collinear data]{Model building with collinear data}

If we explore the local condition numbers for models with different structures, it may be possible to build GW regression models which avoid collinearity. Here, we code a function to calibrate and then estimate a basic (un-adjusted) GW regression.  This function can then be used to assess various forms of the model, where the output each time is a vector of local condition numbers for the model that has been fitted.  This function is presented as follows, together with an example model run.

\begin{CodeChunk}
\begin{CodeInput}
R> test.CN <- function(model, data) {
+   lcrmx.bw <- bw.gwr.lcr(model, data = data, kernel = "bisquare", 
+   adaptive = TRUE)
+   print(model)
+   print(lcrmx.bw)
+   lcrmx    <-    gwr.lcr(model, data = data, bw = lcrmx.bw, 
+   kernel = "bisquare", adaptive = TRUE)
+   print(summary(lcrmx$SDF$Local_CN))
+   lcrmx$SDF$Local_CN
+ }

R> data <- Dub.voter

R> model <- as.formula(GenEl2004 ~ DiffAdd + LARent + SC1 + Unempl + 
+ LowEduc + Age18_24 + Age25_44 + Age45_64)
R> AllD  <- test.CN(model,data)
\end{CodeInput}
\end{CodeChunk}

\begin{figure}
\centering
  \includegraphics[width=1\linewidth]{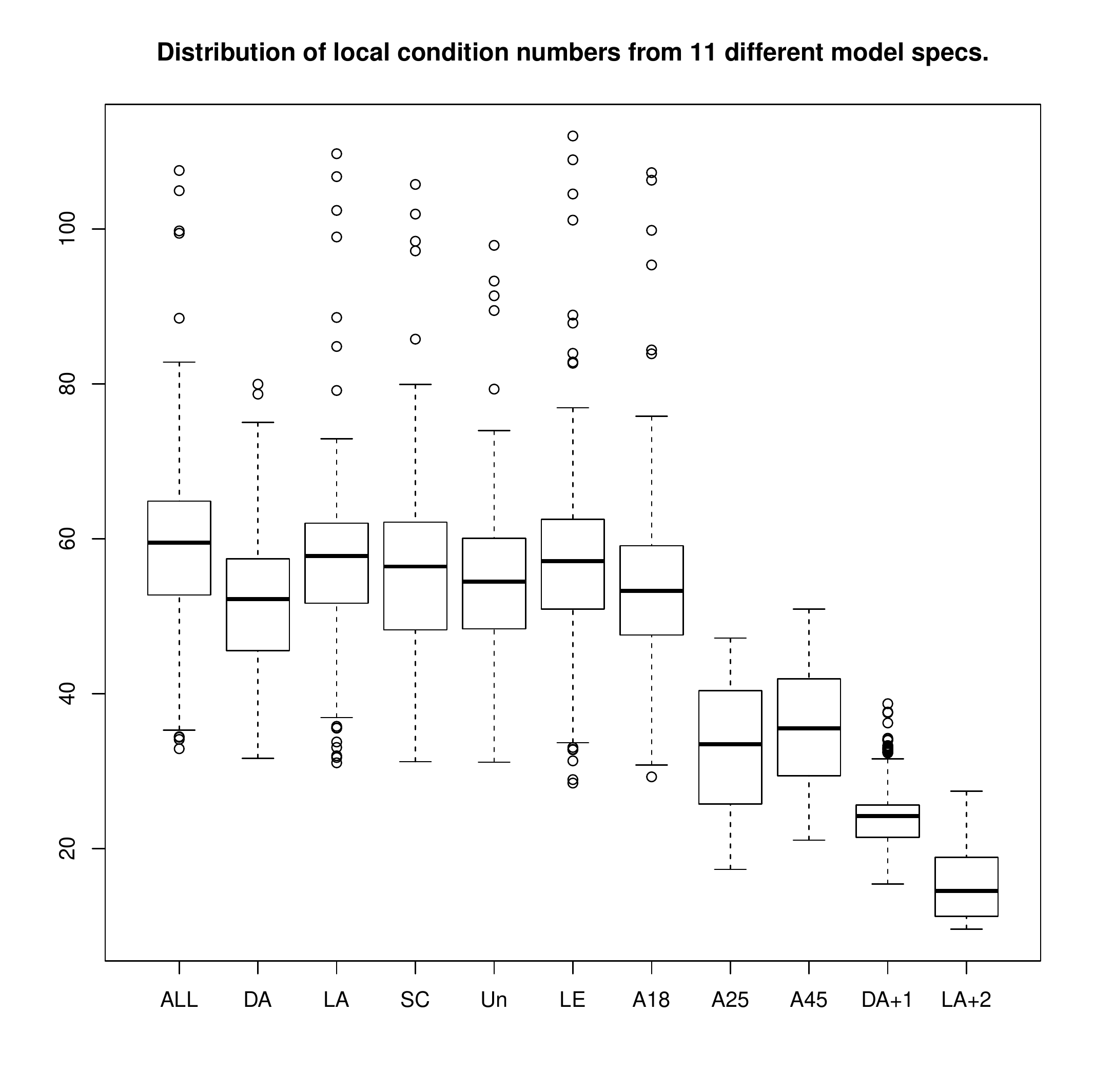}
  \caption{Distribution of local condition numbers from 11 different GW regression fits.}
\label{fig:gwcoll12}
\end{figure}

On using this function with eleven different GW regression models, Figure~\ref{fig:gwcoll12} shows the boxplots of the local condition numbers from GW regressions with: (i) all variables (ALL), (ii) removing each variable in turn (\code{DiffAdd}, \code{LARent}, \code{SC1}, \code{Unempl}, \code{LowEduc}, \code{Age18_24}, \code{Age25_44}, \code{Age45_64}), (iii) removing \code{DiffAdd} and \code{Age45_64} together, and (iv) removing \code{LARent}, \code{Age25_44} and \code{Age45_64} together. The last grouping was suggested by the output of the global PCA from Section~\ref{sec.gwpca}.  Removing variables individually has little effect on the local condition number distributions, although removing the last two age variables induces a noticeable drop. Removing the most collinear variables produces a model where no local condition number is above 30.

\begin{figure}
\centering
  \includegraphics[width=1\linewidth]{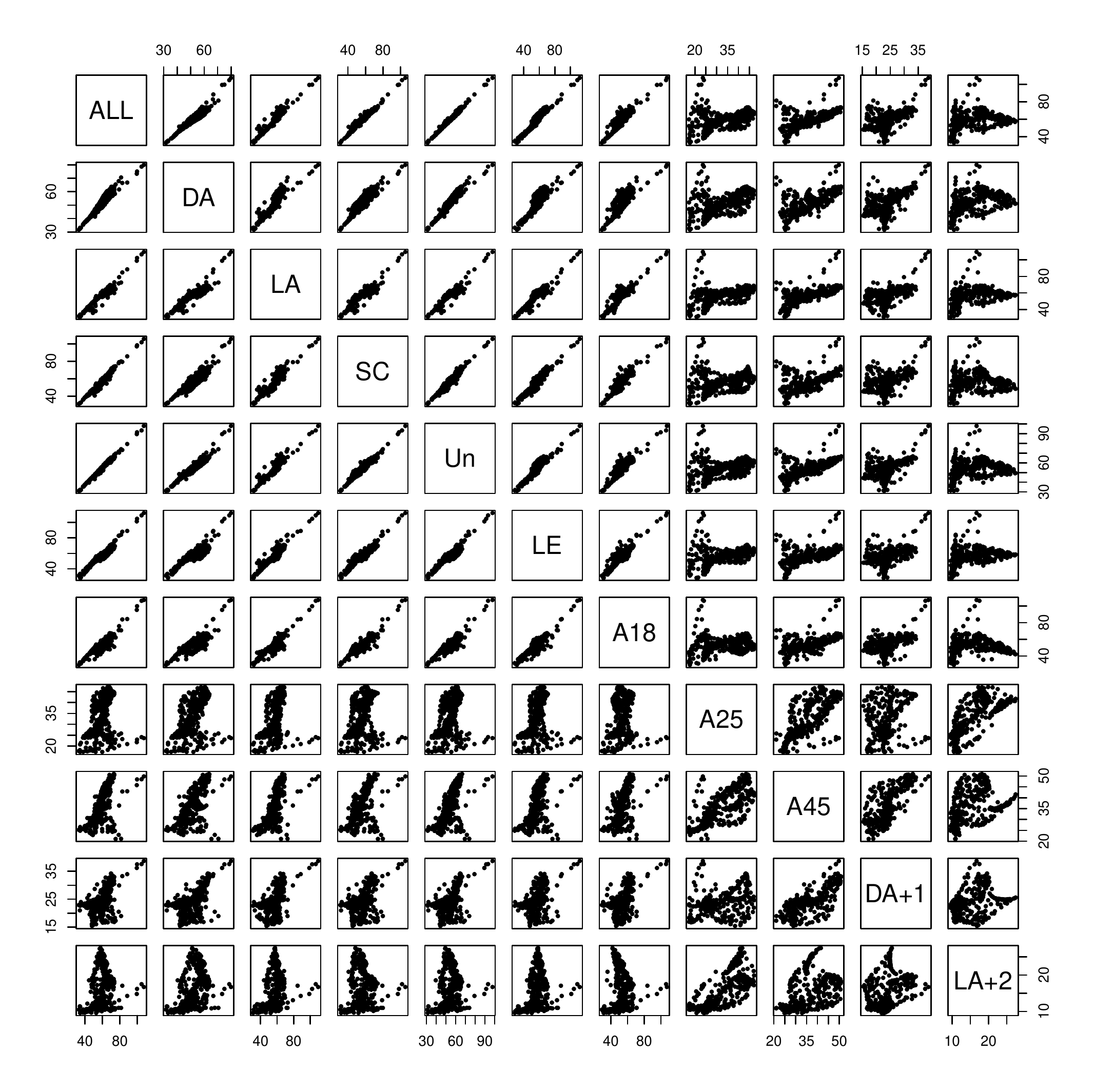}
  \caption{Scatterplot matrix of local condition numbers from 11 different GW regression fits.}
\label{fig:gwcoll13}
\end{figure}

Figure~\ref{fig:gwcoll13} shows the local condition number distributions as a scatterplot matrix.  Here the least effective variable removals have high correlations between the condition number distributions, whereas removing the most collinear variables tends to provide lower condition number correlations with other model forms. This opens up the possibility of semi-automating the model building process to yield a GW regression with acceptably low levels of collinearity.

\subsection[LCR GW regression vs. previous penalised GW regression models]{LCR GW regression vs. previous penalised GW regression models}

It is important to clarify the difference between our LCR GW regression (say, LCR-GWR) and the GW ridge regression (GWRR) demonstrated in \cite{whe07}.  Essentially, LCR-GWR is more locally-focused than GWRR.  GWRR similarly applies a local compensation, but for each local regression, the ridge parameter is global (i.e., it does not vary across space). This global ridge is used everywhere, ensuring that all local regressions of GWRR are biased.  Whereas for LCR-GWR, local ridge parameters are used, and they are only used at locations where they are most needed (as set by a condition number threshold).  At other locations, ridges of zero are specifed.  Thus depending on the condition number threshold set, not all of the local regressions of a LCR-GWR model are necessarily biased. Wheeler actually noted this short-coming of GWRR and suggested the use of local ridge terms \citep[p.2480]{whe07}, but provided no details on such a model's implemention. Thus the advance for LCR-GWR is the actual implementation of a such a model, and in particular, the relating of the local condition number to the local ridge parameter (Section~\ref{sec.lcrgwr}).

LCR-GWR should not only be placed in context with GWRR, but also the GW lasso (GWL).  Ridge regression and the lasso \citep{tib96} both penalise or shrink regression coefficients, so as to reduce collinearity effects, where the lasso shrinks the least significant coefficents to zero, and in doing so, additionally (and simultaneously) provides a model selection function. \cite{whe09} provides two versions of GWL; one that estimates a single (global) lasso parameter to control coefficient shrinkage (GWL-global), and one that estimates multiple (local) lasso parameters (GWL-local).  Thus GWL-global can be viewed as an alternative to GWRR (as both are globally-focused), whilst GWL-local can be viewed as an alternative to LCR-GWR (as both are locally-focused). 

A simple way to demonstrate the differences between the competing models is to run them (togther with a basic GW regression), using the same data and the same kernel weighting specification.  Here the functions, \code{gwrr.est} and \code{gwl.est} from the \pkg{gwrr} \proglang{R} package are needed to fit the GWRR and GWL models, respectively. Correspondence with the author of \pkg{gwrr} indicated that \code{gwl.est} only fits GWL-local models, and that an option for GWL-global is not provided. Furthermore, \code{gwrr.est}, \code{gwl.est} and \code{gwr.est} (a basic GW regression function in \pkg{gwrr}), only allow for fixed bandwidths with Gaussian or exponential kernel functions (see Section~\ref{sec.gwcomp}). As the bandwidth for GWL-local is optimally estimated (via cross-validation) within the \code{gwl.est} function, and cannot be user-specified, we use this particular bandwidth for all models. Thus, only GWL-local will be correctly specified with respect to its weighting function. For a data set, we choose the \code{columbus} neighbourhood crime rate data available in \pkg{gwrr}.

Commands to conduct this rudimentary comparison are given below, where an exponential kernel bandwidth of 1.68 is used for all models. For GWRR and LCR-GWR, only the latter provides local ridges, 16 of which are non-zero (as their corresponding local condition numbers are above 30). The only outputs that are directly comparable between all four models, are the local regression coefficients themselves. For example, if we view the coefficients for \code{income}, 26 are zero with GWL-local, indicating that this variable is not needed in the corresponding local regressions. If we compare the \code{income} coefficients for the penalised models with those from the basic model, both GWRR and GWL-local apply some form of penalty at all locations, whereas LCR-GWR only penalises at certain (in this case, 16) locations. Thus as would be expected, coefficients from the basic GW regression have the strongest correlation with those from the LCR-GWR model. An objective and detailed comparison of all such penalised models, using simulated data, is a subject of our current research \citep{Coll2012}.

\begin{CodeChunk}
\begin{CodeInput}
R> library("gwrr")

R> data("columbus")

R> locs <- cbind(columbus$x, columbus$y)
R> columbus.spdf <- SpatialPointsDataFrame(locs,columbus)

R> gwl.comp <- gwl.est(crime ~ income + houseval, locs, columbus, "exp")

R> gwl.comp$phi
\end{CodeInput}
\begin{CodeOutput}
[1] 1.678067
\end{CodeOutput}
\begin{CodeInput}
R> summary(gwl.comp$beta[2,])
\end{CodeInput}
\begin{CodeOutput}
   Min. 1st Qu.  Median    Mean 3rd Qu.    Max. 
-2.6690 -1.1400  0.0000 -0.5731  0.0000  0.6526 
\end{CodeOutput}
\begin{CodeInput}
R> lcr.gwr.comp <- gwr.lcr(crime ~ income + houseval, data=columbus.spdf,
+ bw=gwl.comp$phi, kernel="exponential", lambda.adjust=TRUE, cn.thresh=30)

R> summary(lcr.gwr.comp$SDF$Local_CN)
\end{CodeInput}
\begin{CodeOutput}
   Min. 1st Qu.  Median    Mean 3rd Qu.    Max. 
  8.245  18.150  22.000  32.120  39.800 109.000 
\end{CodeOutput}
\begin{CodeInput} 
R> summary(lcr.gwr.comp$SDF$Local_Lambda)
\end{CodeInput}
\begin{CodeOutput}
    Min.  1st Qu.   Median     Mean  3rd Qu.     Max. 
0.000000 0.000000 0.000000 0.008032 0.014580 0.042930  
\end{CodeOutput}
\begin{CodeInput}
R> gwrr.comp <- gwrr.est(crime ~ income + houseval, locs,
+ columbus, "exp", bw=gwl.comp$phi)

R> gwrr.comp$lambda
\end{CodeInput}
\begin{CodeOutput}
[1] TRUE
\end{CodeOutput}
\begin{CodeInput}
R> gwr.comp <- gwr.est(crime ~ income + houseval, locs,
+ columbus, "exp", bw=gwl.comp$phi)

R> cor(cbind(gwl.comp$beta[2,], lcr.gwr.comp$SDF$income,
+ gwrr.comp$beta[2,], gwr.comp$beta[2,]))
\end{CodeInput}
\begin{CodeOutput}
          [,1]      [,2]      [,3]      [,4]
[1,] 1.0000000 0.5923885 0.2228593 0.6160248
[2,] 0.5923885 1.0000000 0.2165015 0.9187365
[3,] 0.2228593 0.2165015 1.0000000 0.3579617
[4,] 0.6160248 0.9187365 0.3579617 1.0000000
\end{CodeOutput}
\end{CodeChunk}   

\subsection[Concluding remarks: criticisms and Bayesian models]{Concluding remarks: criticisms and Bayesian models}

Collinearity is a problem in any form of regression modelling and the importance of assessing and taking action has been raised by the many commentaries on local collinearity in GW regression \citep{WheTie:05a, whe07, whe09, whe10, whe13, gri08, PaFaWe11}. Collinear data require care and the tools available in \pkg{GWmodel} should help the analyst assess the magnitude of the problem, where it is a problem, and lead the analyst to take appropriate action.

Properly accounting for the effects of collinearity in GW regression is one of three recommendations for the practical application of GW regression, given in the simulation study of \cite{PaFaWe11}. A second and similary important recommendation is that the bandwidth should be chosen with care.  In our view, if an optimally found bandwidth tends to its maximum, then the underlying process should be viewed as stationary.  Optimal bandwidths that are very small should also be viewed suspiciously, suggesting a under-lying random process.  In this respect, \pkg{GWmodel} provides additional functions to more fully interrogate bandwidth choice, complementing the automated procedures (see Section~\ref{sec.disc}). It should always be noted that an optimal bandwidth for GW regression is based on the accurate prediction of the response variable, not (as it ideally should be) the accurate prediction of the coefficients (which afterall, are the ones used to assess relationship non-stationarity).  A further consideration is that bandwidth size will directly effect the nature of local collinearity. In this respect, \cite{Coll2012} propose an alternative locally-compensated GW regression, where the bandwidth is increased only at locations most seriously effected by collinearity.  In essence, the bandwidth selection function in any GW model should always be thoroughly investigated, and can be considered analogous to a thorough investigation of the variogram in geostatistics; where both investigations aim to identify spatial structure in some way \citep[see also][]{cre89}.      

A third recommendation of \cite{PaFaWe11} concerns sample size, where it is not recommended to use GW regression for data sets of $n<160$.  In our view, this threshold could be set lower, but provided a thorough and thoughtful GW regression analysis is undertaken, using many of the tools described here.  As indicated by \cite{PaFaWe11}, the interpretation of the results should "proceed with extreme caution", in such cases.  Observe that this third recommendation can be just as applicable to any non-stationary spatial model, not just GW regression. Considering all three recommendations together, ensures that GW regression should never be viewed as a 'black box' technique.

Once a thorough GW regression analysis has been conducted, the exploratory and hopefully, reliable results may need to be extended for statistical inference.  Here, GW regression does not provide a coherent inferential framework, since no one model exists.  Instead GW regression yields a collection of models, calibrated with the same data, but with different weighting schemes.  In this respect, any hypothesis testing with GW regression (for example an investigation of the local (pseudo) \emph{t}-values) should only ever be viewed as explorative.

To this extent, it may be worthwhile to consider a Bayesian spatially varying coefficient (SVC) model \citep{gelf03, ass03, whecald07, wal07, whewal09, fin11}. These non-stationary regressions do provide a coherent inferential framework (as a single model exists) and re-assuringly, do not suffer from the local collinearity problems found in GW regression.  Bayesian SVC models do however suffer from their computational complexity, where often only the simplest models can be run (e.g. with only two or three predictors).  Furthermore, software for their implementation is often not readily available \citep [p.153]{fin11}.  Until these modelling issues are resolved, it is likely that GW regression will continue to provide an important exploratory technique for the investigation of relationship non-stationarity; which on occasion, may help direct and parameterise, a subsequent Bayesian SVC model fit and analysis.

\section[GW regression for spatial prediction]{GW regression for spatial prediction}\label{sec.gwpred}
The use of GW regression as a spatial predictor has attracted much attention \citep[e.g., the empirical evaluations of][]{gaasch06,PaFeFa08,lloy:10b}, where it has often performed well, relative to a geostatistical (kriging) alterative.  More objective, simulation experiments can be found in \cite{hafocrch10}, where kriging performed the best, but GW regression still performed with merit.  Conceptual links between GW regression and kriging are discussed in \cite{harbrfo11} and various hybrid predictors can be found in \cite{lloy:10b,hafocrch10,haju11}, where GW regression can provide a useful trend component of a kriging model.

Thus studies have demonstrated value in using GW regression as a spatial predictor.  This is not surprising given that GW regression and associated tools,  directly stem from the non-parametric, locally weighted regression models of \cite{cleve79, clevedev88}, for curve fitting and interpolation.  Furthermore, when used as spatial predictor, the many criticisms levelled at GW regression, such as problems with collinearity (Section~\ref{sec.multicoll}), are not expected to be of such importance. In this respect, our current research is investigating this topic, via simulated data experiments.

To use GW regression as a spatial predictor is straight-forward, a GW regression prediction at a location $\mathrm{s}$ can be found using:
\begin{equation*}
\hat{y}_{GWR}(\mathrm{s})=x(\mathrm{s})^\top \hat{\beta}(\mathrm{s})
\end{equation*}
where $x(\mathrm{s})$ and $\hat{\beta}(\mathrm{s})$ are each   vectors of independent data values and parameter estimates, respectively.  Following \cite{LeMeZh:00a}, the corresponding GW regression prediction variance at $\mathrm{s}$ can be estimated using:
\begin{equation*}
\sigma^2_{GWR}(\mathrm{s})=\mathrm{VAR}\left\{\hat{y}(\mathrm{s})-y(\mathrm{s})\right\}=\hat{\sigma}^2\left[1+\mathrm{S}(\mathrm{s})\right]
\end{equation*}
where
\begin{equation*}
\mathrm{S}(\mathrm{s})=x(\mathrm{s})^\top \left[X^\top W(\mathrm{s})X\right]^{-1}X^\top W^2(\mathrm{s})X\left[X^\top W(\mathrm{s})X\right]^{-1}x(\mathrm{s})
\end{equation*}
Here an unbiased estimate of the residual variance is taken as $\hat{\sigma}=\mathrm{RSS}/(n-\mathrm{ENP})$, where RSS is the residual sum of squares and ENP is the effective number of parameters of the GW regression model.  Observe that the prediction variance for GW regression is for a single observation and not for the mean, and in doing so, is directly comparable with that found with kriging, in say, the \pkg{gstat} \proglang{R} package \citep{peb04}.

\subsection[Example]{Example}
\begin{figure}[!ht]
\centering
\begin{subfigure}{.5\textwidth}
  \centering
  \includegraphics[width=1\linewidth]{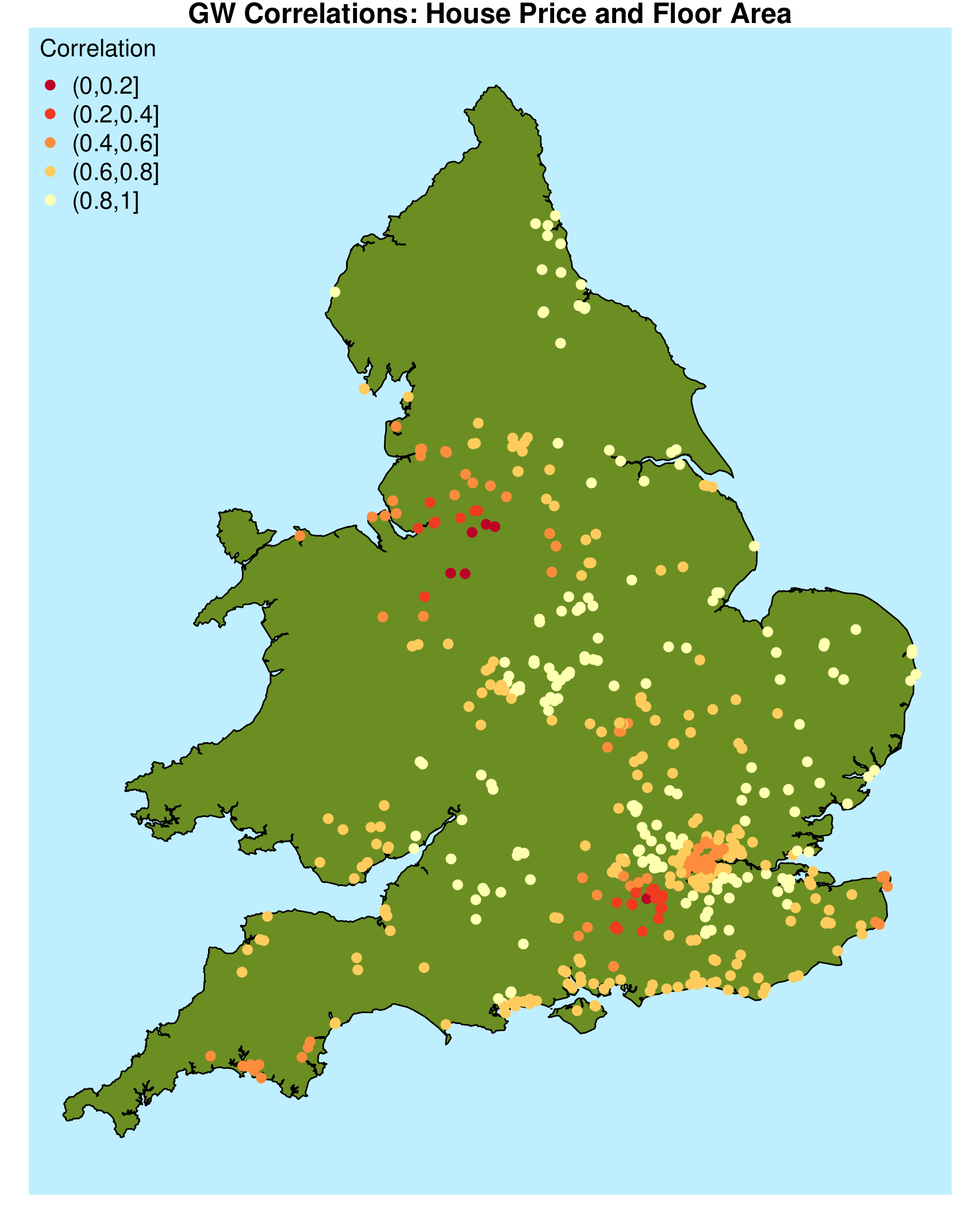}
  \caption{}
  \label{fig:gwpred_14a}
\end{subfigure}%
\begin{subfigure}{.5\textwidth}
  \centering
  \includegraphics[width=1\linewidth]{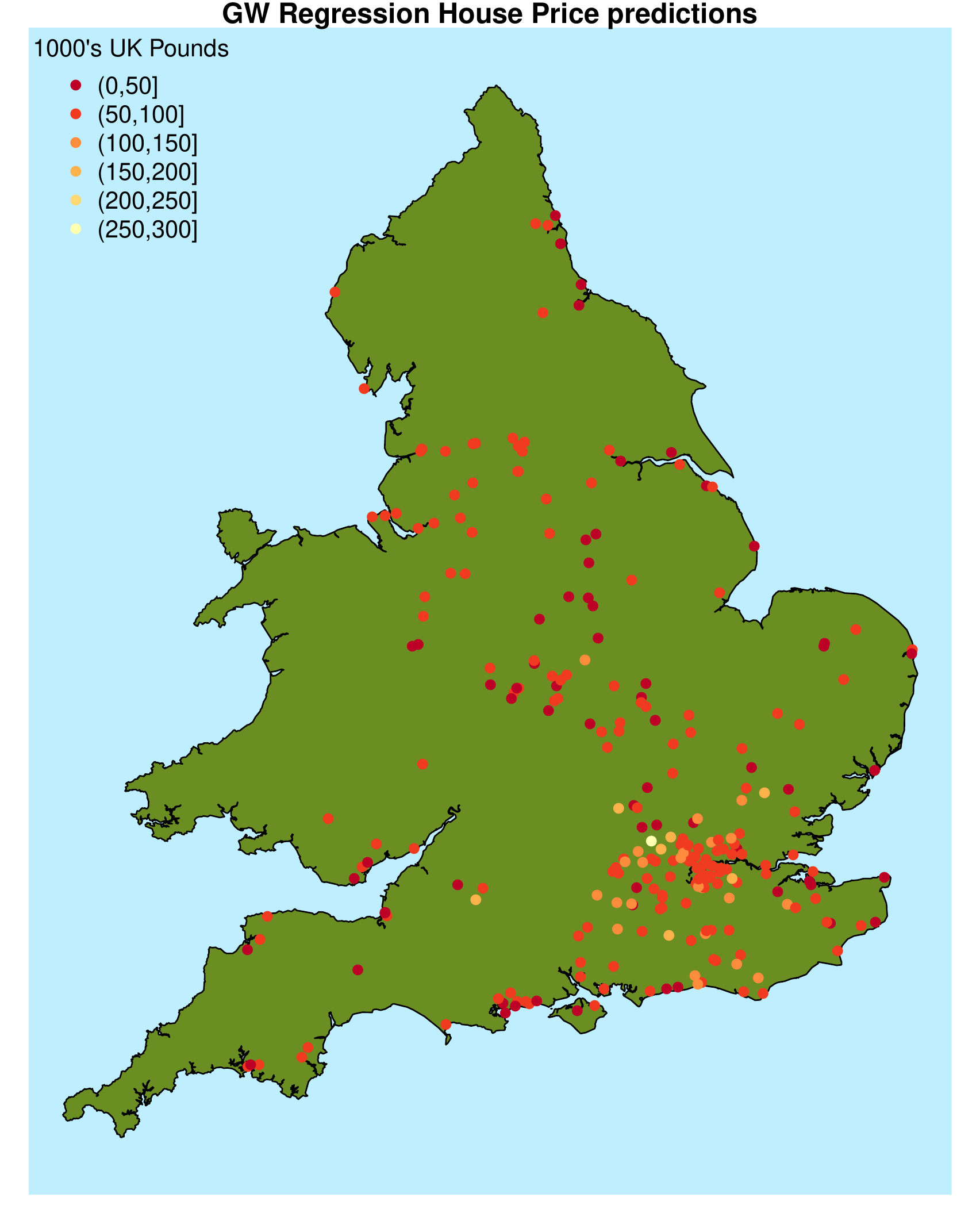}
  \caption{}
  \label{fig:gwpred_14b}
\end{subfigure}
\caption{(a) GW correlations between house price and floor area and (b) GW regression predictions of house price.}
\label{fig:gwpred14}
\end{figure}

To demonstrate GW regression as spatial predictor, we use the \code{EWHP} data set.  Here our aim is to predict the dependent variable, house price (\code{PurPrice}) using a subset of the nine independent variables described in Section~\ref{sec.dataset}, each of which reflect some hedonic characteristic of the property.  A correlation analysis reveals that \code{FlrArea} (floor area or the effective size of the property) provides the strongest correlation with house price at $\rho=0.65$.  For demonstration purposes, we choose to focus our prediction models using only this single hedonic variable.  Two additional \pkg{R} libraries are required for this analysis:
\begin{CodeChunk}
\begin{CodeInput}
R> library("ModelMap")
R> library("gstat")
\end{CodeInput}
\end{CodeChunk}
A GW correlation analysis between house price and our retained hedonic variable\footnote{Observe that discarding hedonic variables that correlate weakly at the global scale does not directly entail similarly weak correlations, locally.  As such, a comprehensive analysis would have locally investigated all relationships with their respective GW correlations (see Section~\ref{sec.gwss}).}  can be conducted using the \code{gwss} function, where GW correlations (and other GW summary statistics) are specified using an adaptive bi-square kernel with a bandwidth of $N=52$ (approximately 10\% of the sample data).  The \code{quick.map} function allows the GW correlations to be mapped.  Figure~\ref{fig:gwpred_14a} displays the resultant map, where the relationship between house price and floor area tends to weaken in rural areas.  This is expected, as large properties are more likely to sell for a premium in urban areas (with the notable exception of some properties located in central London).  Commands for this analysis are as follows:
\begin{CodeChunk}
\begin{CodeInput}
R> ewhp.spdf <- SpatialPointsDataFrame(ewhp[,1:2], ewhp)
R> data("EWOutline")

R> gw.sum.stats <- gwss(ewhp.spdf, vars = c("PurPrice", "FlrArea"),
+ kernel = "bisquare", adaptive = TRUE, bw = 52)
R> quick.map <- function(spdf, var, legend.title, main.title) {
+  x <- spdf@data[,var]
+  cut.vals <- pretty(x)
+  x.cut <- cut(x, cut.vals)
+  cut.levels <- levels(x.cut)
+  cut.band <- match(x.cut, cut.levels)
+  colors <- rev(brewer.pal(length(cut.levels), 'YlOrRd'))
+  par(mar = c(1, 1, 1, 1))
+  plot(ewoutline, col = 'olivedrab', bg = 'lightblue1')
+  title(main.title)
+  plot(spdf, add = TRUE, col = colors[cut.band], pch = 16)
+  legend('topleft', cut.levels, col = colors, pch = 16, bty = 'n',
+  title = legend.title)
+ }

R> X11(width = 8, height = 10)
R> quick.map(gw.sum.stats$SDF, "Corr_PurPrice.FlrArea",
+ "Correlation", "GW Correlations: House Price and Floor Area")
\end{CodeInput}
\end{CodeChunk}

GW correlations provide evidence of non-stationarity in a specific house price relationship and as such, the use of GW regression to predict house price is worth pursuing.  To this extent, we compare the prediction accuracy of a GW regression fit with that of its corresponding (ordinary least squares) global regression fit\footnote{A more complete analysis would also calibrate geostatistical/autocorrelation-based predictors for comparison (e.g., \cite{PaFeFa08}, with respect to house price prediction).}.  Here we split the EWHP data into model calibration and model validation data sets of equal size using the function \code{get.test} in the \pkg{ModelMap} \pkg{R} package.  To find an optimal bandwidth for the GW regression, the function \code{bw.gwr} is used with the calibration data and an optimal adaptive bandwidth of $N=34$ results (for a bi-square kernel, via the CV approach).  We then parameterise the function \code{gwr.predict} with this bandwidth, to find the GW regression predictions and prediction variances at the validation sites.  The necessary distance matrices are found using the function \code{gw.dist}.  The corresponding global regression is also found using \code{gwr.predict}; and as a check for consistency, similarly found using functions \code{gstat} and \code{predict} from the \pkg{gstat} package. The commands used are as follows:

\begin{CodeChunk}
\begin{CodeInput}
R> write.table(ewhp, "ewhp.csv", col.names = T, row.names = F, sep=",")
R> get.test(proportion.test = 0.5, "ewhp.csv", seed = 42, folder = getwd(),
+ qdata.trainfn = "ewhp_calib.csv", qdata.testfn = "ewhp_valid.csv")

R> ewhp_calib <- read.table("ewhp_calib.csv", header = T,sep = ",")
R> attach(ewhp_calib)
R> ewhp_calib.spdf <- SpatialPointsDataFrame(ewhp_calib[,1:2],
+ as.data.frame(ewhp_calib[c(3,12)]))

R> dm.calib <- gw.dist(dp.locat = coordinates(ewhp_calib.spdf))

R> gwr.bw.cv <-bw.gwr(PurPrice ~ FlrArea,data = ewhp_calib.spdf,
+ approach = "CV", kernel = "bisquare", adaptive = T, dMat = dm.calib) 

R> ewhp_valid <- read.table("ewhp_valid.csv", header = T, sep = ",")
R> attach(ewhp_valid)
R> ewhp_valid.spdf <- SpatialPointsDataFrame(ewhp_valid[,1:2],
+ as.data.frame(ewhp_valid[c(3,12)]))

R> dm.valid <- gw.dist(dp.locat = coordinates(ewhp_calib.spdf),
+ rp.locat = coordinates(ewhp_valid.spdf))

R> gwr.pred <- gwr.predict(PurPrice ~ FlrArea,data = ewhp_calib.spdf,
+ predictdata = ewhp_valid.spdf ,bw = gwr.bw.cv, kernel = "bisquare",
+ adaptive = T, dMat1 = dm.valid, dMat2 = dm.calib)

R> ols.pred.gwmodel <- gwr.predict(PurPrice ~ FlrArea, data = ewhp_calib.spdf,
+ predictdata = ewhp_valid.spdf, bw = 519, kernel = "boxcar", adaptive = T,
+ dMat1 = dm.valid, dMat2 = dm.calib)

R> ols <- gstat(id = "mlr",formula = PurPrice ~ FlrArea,
+ loc=~Easting + Northing, data = ewhp_calib)
R> ols.pred.gstat <- predict(ols,newdata = ewhp_valid,BLUE = TRUE)
\end{CodeInput}
\end{CodeChunk}

Performance results are reported in Table~\ref{tab.1ch8}, in terms of prediction accuracy and prediction uncertainty accuracy.  Prediction accuracy is measured by the root mean squared prediction error (RMSPE) and the mean absolute prediction error (MAPE), both of which should tend to zero.  Prediction uncertainty accuracy is measured by the mean and standard deviation (SD) of the prediction z-score data (mean.ZS and SD.ZS, respectively).  These z-scores are defined as:
\begin{equation*}
\mathrm{z-score}_{pred}(\mathrm{s})=\left(y(\mathrm{s})-\hat{y}(\mathrm{s})\right)/\sigma_{pred}(\mathrm{s})
\end{equation*}
where for unbiased prediction standard errors, the mean and SD of the z-scores should tend to zero and unity, respectively.  As would be expected, GW regression provides the best set of results and there is near exact correspondence between the two global regression results.  Finally, a map depicting the GW regression predictions is given in Figure~\ref{fig:gwpred_14b}.  Commands to conduct this model performance analysis, just for the GW regression model, are as follows:

\begin{CodeChunk}
\begin{CodeInput}
R> RMSPE.gwr <- 
+ (mean((ewhp_valid.spdf$PurPrice - gwr.pred$SDF$prediction)^2))^0.5
R> MAPE.gwr <- mean(abs(ewhp_valid.spdf$PurPrice - gwr.pred$SDF$prediction))
R> zscore.gwr <- (ewhp_valid.spdf$PurPrice - gwr.pred$SDF$prediction)/
+ (gwr.pred$SDF$prediction_var)^0.5
R> MeanZ.gwr <- mean(zscore.gwr)
R> SDZ.gwr <- (var(zscore.gwr))^0.5

R> gwr.pred$SDF$prediction <- gwr.pred$SDF$prediction/1000

R> X11(width = 8, height = 10)
R> quick.map(gwr.pred$SDF, "prediction", "1000's UK Pounds",
+  "GW Regression House Price predictions")
\end{CodeInput}
\end{CodeChunk}

\begin{table}
\caption{Performance results for GW Regression as a spatial predictor.}
\begin{tabular}{ l | c | c | c | c }
  \hline   
  Model	& RMSPE (x\pounds1000) & MAPE (x\pounds1000) &	Mean.ZS & SD.ZS\\
  \hline                   
GW Regression	& 27.03 &	17.55 &	-0.04 &	1.10\\
Global Regression (\pkg{GWmodel}) &	31.58 &	20.65 &	0.16 &	1.10\\
Global Regression (\pkg{gstat}) &	31.58 &	20.65 &	0.16 &	1.10 \\
  \hline  
\end{tabular} \label{tab.1ch8}
\end{table}

\section[Comparisons with spgwr, gwrr and McSpatial]{Comparisons with  \pkg{spgwr},  \pkg{gwrr} and  \pkg{McSpatial}}\label{sec.gwcomp}
In this section, we compare the functions of \pkg{GWmodel} with those found in the related packages of \pkg{spgwr} (Version 0.6-24), \pkg{gwrr} (Version 0.2-1) and \pkg{McSpatial} (Version 2.0). Here we establish where they duplicate (or have strong similarities with) each other and where they diverge. This comparison is given as a series of tables. Weighting matrix specifications are summarised in Table~\ref{tab.1ch9}. Functions for GW summary statistics are summarised in Table~\ref{tab.2ch9}. Functions for basic GW regression are summarised in Table~\ref{tab.3ch9}. Functions for generalised GW regression are summarised in Table~\ref{tab.4ch9}. Functions for other GW regressions are given in Table~\ref{tab.5ch9}. Functions for addressing collinearity issues in GW regressions are given in Table~\ref{tab.6ch9}. Lastly, functions for other GW models (including a GW PCA and a GW discriminant analysis) are given in Table~\ref{tab.7ch9}. Clearly, \pkg{GWmodel} provides a more extensive set of GW models and associated tools than any existing package.

\begin{table}[!ht]
\caption{Weighting matrix specifications.}
\begin{tabular}{ l | c | c | c | c }
  \hline   
  	& \pkg{GWmodel} & \pkg{spgwr} &	\pkg{gwrr} & \pkg{McSpatial}\\
  \hline                   
Kernel 	& Box-car, &	Bi-square, &	Gaussian &	Rectangular,\\
functions:   & Bi-square, &	Tri-cube &	and Exponential &	Triangular,\\
	& Tri-cube, &	and Gaussian &  & Epanechnikov,\\
	& Gaussian & & & Bi-square,\\
	& and Exponential &	 &  & Tri-cube, \\
	& & & &	Tri-weight\\
	& & & &	and Gaussian\\
  \hline
Adaptive &	Yes &	Yes with &	No &	No\\
  bandwidth? &	 &	\code{gwr.adapt}  &	&	\\
  \hline
Fixed &	Yes &	Yes &	Yes &	Yes \\
bandwidth? &	& & &  \\
  \hline
Spatial &	Euclidean, &	Euclidean &	Euclidean &	Euclidean \\
distance  & Great Circle 	 &	and Great Circle   &	& and Great Circle 	\\
 metrics: &	and Minkowski &	 &	&	\\
  \hline
Functions &	Yes with &	Yes with &	No &	Yes with \\
for weights &	\code{gw.dist} &	\code{gwr.bisquare}, &	 &	\code{makew} \\
matrix  &  &	\code{gwr.tricube} & & \\
computation?  &  &	and \code{gwr.gauss} & & \\
  \hline  
\end{tabular} \label{tab.1ch9}
\end{table}

\begin{table}[!ht]
\caption{GW summary statistics functions.}
\begin{tabular}{ l | c | c | c | c }
  \hline   
  	& \pkg{GWmodel} & \pkg{spgwr} &	\pkg{gwrr} & \pkg{McSpatial}\\
  \hline                   
Basic?	& Yes with \code{gwss}&	Yes with \code{gw.cov} &	No &	No\\
     \hline
Basic & mean & mean &   & \\
statistics:	& standard deviation & standard deviation & & \\
	& variance & std. error of mean	 &  & \\
	& skew & std. difference of global mean	 &  & \\
	& coeff. of variation & std. difference of local mean	 &  & \\
	& covariance & covariance	 &  & \\
	& Pearson's correlation & Pearson's correlation	 &  & \\
  \hline
Robust?	& Yes with \code{gwss}&	No &	No &	No\\
  \hline
Robust & median &  &   & \\
statistics:	& inter-quartile range & & & \\
	& quantile imbalance & & & \\
	& Spearman's rank correl. & & & \\
	  \hline
Monte-Carlo & Yes with &	No &	No &	No\\
tests? & \code{montecarlo.gwss} &	 &	 & \\
	\hline 
\end{tabular} \label{tab.2ch9}
\end{table}

\begin{table}[!ht]
\caption{Basic GW regression functions.}
\begin{tabular}{ l | c | c | c | c }
  \hline   
 	& \pkg{GWmodel} & \pkg{spgwr} &	\pkg{gwrr} & \pkg{McSpatial}\\
  \hline                   
Basic	& Yes with &	Yes with &	Yes with &	Yes with\\
 GW regression?   & \code{gwr.basic}  &	\code{gwr} & \code{gwr.est}	 & \code{cparlwr}\\
     \hline
CV	& Yes with &	Yes with &	Yes with &	Yes with\\
bandwidth?  & \code{bw.gwr}  &	\code{gwr.sel} & \code{gwr.bw.est}	 & \code{cparlwrgrid}\\
   \hline
Generalised CV 	& No &	No &	No &	Yes with\\
bandwidth?   &  &  & 	 & \code{cparlwrgrid}\\
   \hline
AICc & Yes with &	Yes with &	No &	No\\
  bandwidth? &\code{bw.gwr}  &	\code{gwr.sel} & 	 & \\
   \hline
Investigative tools	& Yes with \code{gwr.cv}&	No &	No &	No\\
for CV bandwidth?   & and \code{gwr.cv.contrib} &	 & 	 & \\
   \hline
Model selection & Yes with &	No &	No &	No\\
 tools?   & \code{model.selection.gwr}, &	 & 	 & \\
    & \code{model.sort.gwr} &	 & 	 & \\
    & and \code{model.view.gwr} &	 & 	 & \\
    \hline
Local vs. global	& Yes with &	Yes with &	No &	No\\
regression tests?   & \code{gwr.basic} & \code{LMZ.F3GWR.test}	 & 	 & \\
   \hline
Monte-Carlo test?	& Yes with &	No &	No &	No\\
   &\code{montecarlo.gwr} & & 	 & \\
   \hline
Multiple hypothesis	& Yes with &	No &	No &	No\\
tests?   &\code{gwr.t.adjust} & & 	 & \\
   \hline
Moran's I test?	& No &	Yes with &	No &	No\\
   & & \code{gwr.morantest}& 	 & \\
   \hline
Prediction variances	& Yes with &	Yes with &	No &	No\\
(or std errors)?   &\code{gwr.predict} & \code{gwr}	 & 	 & \\
   \hline
Parallel computing	& No &	Yes with &	No &	No\\
options?   & & \code{gwr}& 	 & \\
   \hline
\end{tabular} \label{tab.3ch9}
\end{table}

\begin{table}[!ht]
\caption{Generalised GW regression functions.}
\begin{tabular}{ l | c | c | c | c }
  \hline   
 	& \pkg{GWmodel} & \pkg{spgwr} &	\pkg{gwrr} & \pkg{McSpatial}\\
  \hline                   
Generalised 	& Yes with  &	Yes with \code{ggwr}&	No &	Yes with \code{cparlogit},\\
GW regression?   &  \code{gwr.generalised}&	 &	 & \code{cparmlogit}\\
   &  &	 &	 & and \code{cparprobit}\\
     \hline
Family/link:	& binomial&	As for \code{glm}&	No & logit, multinomial\\
   & and poisson &	 & 	 & and probit\\
   \hline
CV bandwidth?	& Yes with \code{bw.ggwr}&	Yes with \code{ggwr.sel}&	No &	Yes with\\
   &  &	 & 	 & \code{cparlwrgrid}\\
   \hline
Generalised CV 	& No &	No &	No &	Yes with\\
bandwidth?   &  &  & 	 & \code{cparlwrgrid}\\
   \hline
AICc bandwidth?	& Yes with \code{bw.ggwr}&	No &	No &	No\\
   \hline
Investigative tools	& Yes with \code{ggwr.cv}&	No &	No &	No\\
for CV bandwidth?   & and \code{ggwr.cv.contrib} &	 & 	 & \\
   \hline
\end{tabular} \label{tab.4ch9}
\end{table}

\begin{table}[!ht]
\caption{Other GW regression functions.}
\begin{tabular}{ l | c | c | c | c }
  \hline   
 	& \pkg{GWmodel} & \pkg{spgwr} &	\pkg{gwrr} & \pkg{McSpatial}\\
  \hline                   
Robust GW regression? 	& Yes with  \code{gwr.robust}&	No &	No &	No\\
     \hline
Heteroskedastic GW regression? 	& Yes with  \code{gwr.hetero}&	No &	No &	No\\
     \hline
Mixed (semipara.) GW regression? 	& Yes with  \code{gwr.mixed}&	No &	No &	Yes with  \code{semip}\\
     \hline
Quantile GW regression? 	& No &	No &	No &	Yes with  \code{qregcpar}\\
     \hline
\end{tabular} \label{tab.5ch9}
\end{table}

\begin{table}[!ht]
\caption{Functions for addressing collinearity issues in GW regression.}
\begin{tabular}{ l | c | c | c | c }
  \hline   
 	& \pkg{GWmodel} & \pkg{spgwr} &	\pkg{gwrr} & \pkg{McSpatial}\\
  \hline                   
Local  & Yes with  \code{gwss} or& Yes with  \code{gw.cov}&	No &	No\\
correlations? &  \code{gwr.collin.diagno}& & &	\\
	  \hline
Local VIFs? 	& Yes with & No &	No &	No\\
	&  \code{gwr.collin.diagno}& & &	\\
     \hline
Local VDPs? 	& Yes with & No & Yes with  \code{gwr.vdp}&	No\\
	&  \code{gwr.collin.diagno}& & &	\\
     \hline
Local condition 	& Yes with \code{gwr.lcr}& No & Yes with  \code{gwr.vdp}&	No\\
numbers?	& or \code{gwr.collin.diagno}& & &	\\
     \hline
GW regression 	& Yes with \code{gwr.lcr}, & No & Yes with  \code{gwrr.est}&	No\\
with a global	& \code{bw.gwr.lcr}& & &	\\
ridge term	& \code{gwr.lcr.cv}& & &	\\
(user-specified)?	& \code{gwr.lcr.cv.contrib}& & &	\\
     \hline
GW regression 	& No & No & Yes with  \code{gwrr.est}&	No\\
with a global	& & & &	\\
ridge term	& & & &	\\
(estimated)?	& & & &	\\
     \hline
GW regression 	& Yes with \code{gwr.lcr}, & No & No&	No\\
with local	& \code{bw.gwr.lcr}& & &	\\
ridge terms?	& \code{gwr.lcr.cv}& & &	\\
	& \code{gwr.lcr.cv.contrib}& & &	\\
     \hline
GW lasso? 	& No & No & Yes with  \code{gwl.est}&	No\\
     \hline
\end{tabular} \label{tab.6ch9}
\end{table}

\begin{table}[!ht]
\caption{Other GW model functions.}
\begin{tabular}{ l | c | c | c | c }
  \hline   
 	& \pkg{GWmodel} & \pkg{spgwr} &	\pkg{gwrr} & \pkg{McSpatial}\\
  \hline                   
GW discriminant analysis? 	& Yes with  \code{bw.gwda} and \code{gwda}&	No &	No &	No\\
     \hline
GW PCA? 	& Yes with  \code{bw.gwpca}, \code{gwpca}, \code{gwpca.cv},&	No &	No &	No\\
 	& \code{gwpca.cv.contrib}, \code{glyph.plot},&	&	 &	\\
 	& \code{check.components}, \code{plot.mcsims},&	&	 &	\\
 	& \code{montecarlo.gwpca.1}&	&	 &	\\
 	& and \code{montecarlo.gwpca.2},&	&	 &	\\
     \hline
GW parallel coordinate plot? 	& Yes with  \code{gw.pcplot}&	No &	No &	No\\
     \hline
\end{tabular} \label{tab.7ch9}
\end{table}

\section[Discussion]{Discussion}\label{sec.disc}
In this presentation of the \pkg{GWmodel} \proglang{R} package, we have demonstrated the use of various GW models to investigate and model different aspects of spatial heterogeneity.  We have focused our presentation on basic and robust forms of GW summary statistics, GW PCA and GW regression.  We have also provided important extensions to GW regression with respect to local collinearity issues and with respect to spatial prediction.  However, we have not fully described all that is available in \pkg{GWmodel}.  Key omissions include:
\begin{enumerate}
\item Functions to investigate bandwidth selection, where a cross-validation score can be found for a given bandwidth (\code{ggwr.cv}, \code{gwr.cv},\code{ gwr.lcr.cv}, \code{gwpca.cv}); and associated functions to find which observations contribute the most (and are potentially outlying) to this cross-validation score (\code{ggwr.cv.contrib}, \code{gwr.cv.contrib}, \code{gwr.lcr.cv.contrib}, \code{gwpca.cv.contrib}).
\item Functions implementing Monte-Carlo tests for GW summary statistics (\code{montecarlo.gwss}), GW regression (\code{montecarlo.gwr}) and GW PCA (\code{montecarlo.gwpca.1}, \code{montecarlo.gwpca.2}).  These functions test whether aspects of the GW model are significantly different to those that could be found under the global model, as artefacts of random variation in the data.
\item Functions for fitting generalised GW regression models (\code{gwr.generalised}, \code{ggwr.cv}, \code{ggwr.cv.contrib}, \code{bw.ggwr}).
\item Further functions for visualising GW PCA outputs (\code{glyph.plot}, \code{check.components}) and for locally visualising the original multivariate data (\code{gw.pcplot}).
\item A function for multiple hypothesis tests with GW regression (\code{gwr.t.adjust}).
\item Functions for fitting mixed (\code{gwr.mixed}) and heteroskedastic (\code{gwr.hetero}) GW regression.
\item Functions for conducting a GW discriminant analysis (\code{gwda}, \code{bw.gwda}).
\end{enumerate}

The use of these functions will be presented elsewhere. Future iterations of \pkg{GWmodel} will include many more functions, such as those needed to implement GW boxplots and GW variograms.  The GW modelling paradigm itself continues to evolve, with not only new GW models, such as the robust GW regressions of \cite{zhme11} or the GW quantile regression of \cite{chdeyama12}; but also, novel usages of existing GW models, such as using GW PCA to detect multivariate spatial outliers \citep{HarBrChJuCl14} or using GW PCA to direct sample re-design \citep{HarJuClBrCh14}.

\newpage
\section*{Acknowledgements}\label{sec.ack}
Research presented in this paper was funded by a Strategic Research Cluster grant (07/SRC/I1168) by Science Foundation Ireland under the National Development Plan. The authors gratefully acknowledge this support.

\bibliography{biblioGWmodel}

\end{document}